\documentclass{sig-alternate-10pt}

\pdfoutput=1 
\usepackage{graphicx}
\usepackage{balance}
\usepackage{comment}
\usepackage{subfigure}
\usepackage{booktabs}
\usepackage{amsmath}
\usepackage{amssymb}
\usepackage{amsfonts}
\usepackage{bm}
\usepackage{colortbl}
\usepackage{tabularx}
\usepackage{color}
\usepackage{epsfig}
\usepackage{xspace}
\usepackage{graphicx}    
\usepackage{url}         %
\usepackage{algorithm}
\usepackage{algpseudocode}
\usepackage{placeins}
\usepackage{multirow}
\usepackage{epstopdf}
\usepackage{tabularx}

\newcommand{\bo}[1]{{\color{black}{#1}}}
\newcommand{\Ambuj}[1]{{\color{blue}{Ambuj}}}

\begin{document}

\title{dRTI: Directional Radio Tomographic Imaging}
%
%
%
%
%

\numberofauthors{1} 
%
\author{
\alignauthor{Bo Wei$^\dag$$^\forall$$^\ddag$, Ambuj Varshney$^\S$, Wen Hu$^\ddag$$^\forall$,\\ Neal Patwari$^\varnothing$$^\bigtriangleup$, Thiemo Voigt$^\forall$$^\S$, Chun Tung Chou$^\dag$} \\
\affaddr{$^\dag$ University of New South Wales, Sydney, NSW, Australia} \\
\affaddr{$^\forall$ SICS, Stockholm, Sweden}\\
\affaddr{$^\ddag$ CSIRO Computational Informatics, Brisbane, Queensland, Australia}\\
\affaddr{$^\S$ Uppsala University, Uppsala, Sweden}\\
\affaddr{$^\varnothing$ University of Utah, Salt Lake City, Utah, USA}\\
\affaddr{$^\bigtriangleup$ Xandem Technology, Salt Lake City, Utah, USA}\\
 \email{\{bwei,ctchou\}@cse.unsw.edu.edu$^\dag$, ambuj.varshney@it.uu.se$^\S$, wen.hu@csiro.au$^\ddag$, } 
\email{npatwari@ece.utah.edu$^\varnothing$, thiemo@sics.se$^\forall$}
%
%
\alignauthor
Authors
}


\maketitle
\begin{abstract}
Radio tomographic imaging (RTI) enables \emph{device free} localisation of people and objects in many challenging environments and situations. Its basic principle is to detect the changes in the statistics of some radio quality measurements in order to infer the presence of people and objects in the radio path. However, the localisation accuracy of RTI suffers from complicated radio propagation behaviours such as multipath fading and shadowing. In order to improve RTI localisation accuracy, we propose to use inexpensive and energy efficient electronically switched directional (ESD) antennas to improve the quality of radio link behaviour observations, and therefore, the localisation accuracy of RTI. We implement a directional RTI (dRTI) system to understand how directional antennas can be used to improve RTI localisation accuracy. We also study the impact of the choice of antenna directions on the localisation accuracy of dRTI and propose methods to effectively choose informative antenna directions to improve localisation accuracy while reducing overhead. We evaluate the performance of dRTI in diverse indoor environments and show that dRTI significantly outperforms the existing RTI localisation methods based on omni-directional antennas.
\end{abstract}

%
%
%
\vspace{-2mm}
\section{Introduction}\label{intro}
\label{sec:intro}
This paper explores the use of \bo{directional} antennas to improve the accuracy of radio tomographic imaging (RTI).
RTI uses a network of small inexpensive wireless devices, placed in the periphery of an Area of Interest (AoI), to estimate the locations of people and objects within the AoI~\cite{WilsonRTI:2010,WilsonVRTI:2011}. A distinguishing feature of RTI is that it is \emph{device free}~\cite{Youssef:2007} in the sense that the people and objects to be tracked do not have to wear any special purpose devices on them. This makes RTI significantly less intrusive compared with many other localisation methods which require people/objects to carry a radio device or tag. In addition, RTI can function in many challenging environments. RTI can work in both line-of-sight (LOS) and non-LOS (NLOS) environments. RTI can work with no or poor lighting conditions because it uses radio waves rather than visible light. The wireless devices for RTI can be placed outside the walls to enable them to \emph{see through walls} or smoke to locate people or objects within the walls. Therefore, RTI has
a broad range of applications in emergency response, security (e.g. hostage rescue), health care, and assisted living etc.~\cite{bocca2013radio,Kaltiokalliograndma2012}.

%
%

\begin{figure}[t]
\center

   \subfigure[]{
        \includegraphics[scale = .22]{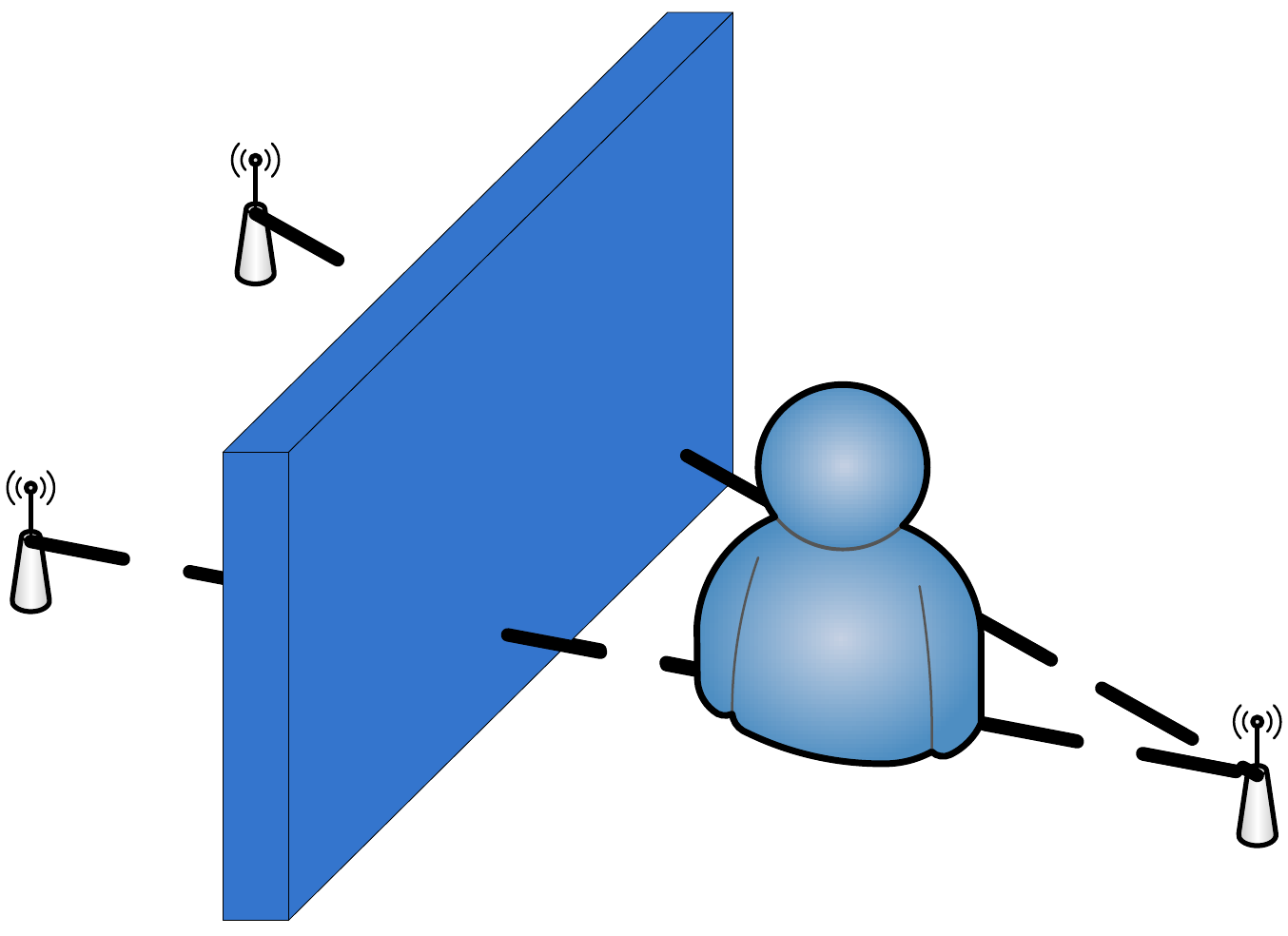}
        \label{fig:example2}
    }
   \subfigure[]{
        \includegraphics[scale = .24]{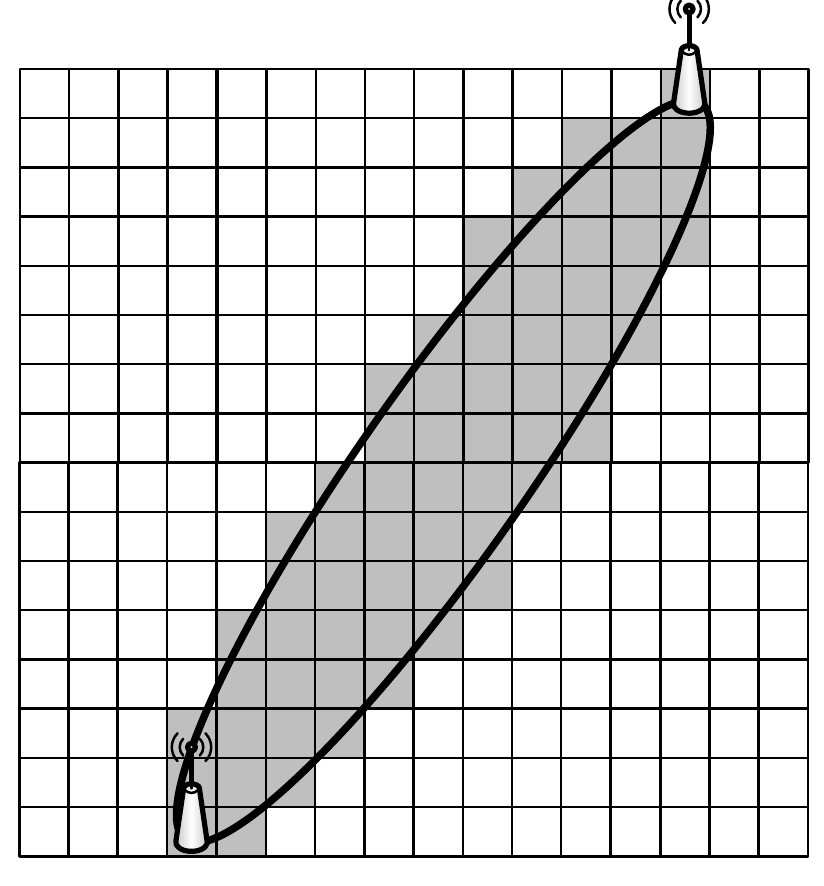}
        \label{fig:example1}
    }
       \subfigure[]{
        \includegraphics[scale = .21]{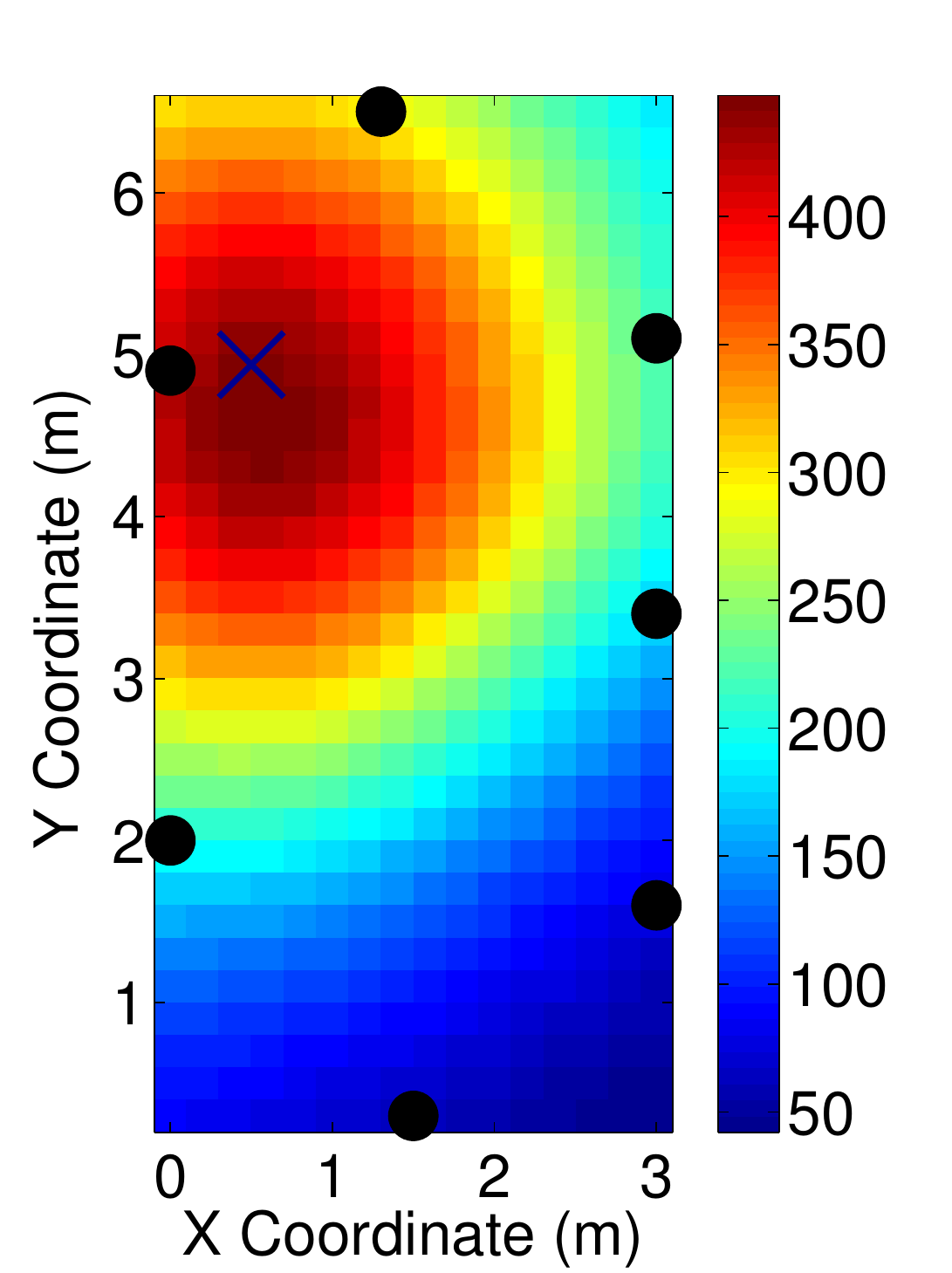}
        \label{fig:example3}
    }

    \caption{(a) A person blocks two radio links, creating a NLOS environment.  (b) The AoI is divided into voxels. An ellipse is used to model the area of attenuation between a pair of nodes.  (c) An example of an RTI image. The blue cross indicates the truth location of the obstruction. (Best view in colour.)}
    \label{fig:example}
    \vspace{-5mm}
\end{figure}


The key idea behind RTI is that people/objects in the AoI present themselves as \emph{obstructions} to radio waves. For example, in Fig.~\ref{fig:example2}, a person stands in the paths of two radio links, and her presence can be detected by observing the \emph{changes} of some statistics of the qualities of the radio links. A more conspicuous change in these radio link quality statistics will result in better location estimate, as well as reduce the number of false positives and negatives. Previous work on RTI relies on omni-directional antennas, which radiate power isotropically in the horizontal plane. Multipath radio channels experience \emph{multipath fading}, in which arriving waves unpredictably experience constructive or destructive interference. Directional antennas reduce the spatial extent of significant multipath, and thus the effects of fading, by focusing the radiated power in some given directions. 
The benefits of RTI with directional antennas (dRTI) are twofold compared to their omni-directional counterparts. First, the presence of obstructions which block the direct paths will have significantly more impact on the link quality behaviour of directional antennas. Second, the 
presence of obstructions which are \emph{outside} the direct paths will have significantly less impact on the link quality of directional antennas. 
Therefore, directional antennas  can provide significantly better link quality observations to improve the RTI localisation accuracy.

Directional antennas have been explored to improve the communication \bo{performance} on mobile phones and wireless networks previously~\cite{AmiriSani:2010:DAD:1859995.1860021, Liu:2009:DII:1592568.1592589}. 
However,
traditional directional antennas such as Yagi antennas have a large form factor and consume a large amount of energy to move from one direction to another direction, which makes them unsuitable for resource-impoverished Wireless Sensor Networks (WSNs) with small nodes. On the other hand,
 Electronically-Switched Directional (ESD) antennas \cite{nilsson2009directional} are significantly more energy efficient compared to transitional directional antennas,
 and enable \emph{dynamic} electronic control of direction of maximum gain. 
Recently, Mottola et al.~\cite{LucaSecon2013} leveraged ESD antennas to alleviate wireless contention in WSNs. In this paper, we, for the first
time, explore the use of ESD antennas to improve RTI localisation accuracy. We detail the feasibility studies for an inexpensive and energy 
efficient dRTI WSN system with ESD antennas and answer the fundamental questions whether and to what extent dRTI
systems can improve tracking performance in practice. 

Although dRTI can significantly improve localisation accuracy, its overhead
 due to the use of directional antenna are high. Consider a transmitter and receiver, each equipped with a directional antenna that yields $n$ different directions, the total number of antenna direction pairs is $n^2$. Hence, conducting link quality measurements on all the possible antenna direction pairs results in higher communication overhead
 and energy consumption. Moreover, not all $n^2$ antenna direction pairs are useful for RTI localisation estimation. We therefore investigate three different methods to select antenna direction pairs to reduce overhead
 while maintaining accurate position estimation. The contributions of this paper are as follows:

1) We conduct comprehensive investigations to demonstrate that directional antennas can produce sharper changes in radio link quality statistics (both mean and variance) compared to their omni-directional counterparts. 

2) We carry out studies to understand how the choice of directional antenna pairs can impact on the radio link quality statistics. Based on these studies, we propose three methods to choose the best directional antenna pairs
to increase the accuracy of RTI while reducing the overhead of the system.

3) We design and implement dRTI, the first system that improves indoor tracking accuracy through the use of ESD antennas, and evaluate the performance of the proposed dRTI system in 
realistic LOS and NLOS environments. Our results show that the proposed dRTI system can improve the localisation 
accuracy significantly compared to state-of-the-art RTI system based on the mean of link quality (mRTI)~\cite{WilsonRTI:2010}, the variance of 
link quality (vRTI)~\cite{WilsonVRTI:2011} and multi-channel RTI (cRTI) using omni-directional antennas.

The rest of this paper is organised as follows. Section~\ref{sec:background} presents background on RTI and ESD antennas. Section ~\ref{sec:methods} presents a few different studies: (1) The impact of obstructions on the link quality statistics for both directional and omni-directional antennas; (2) The variation of link quality statistics for different choices of directional antenna pairs. Based on these studies, we propose three different methods to choose effective directional antenna pairs to reduce overhead
 in dRTI. We present comprehensive evaluations of our dRTI system in Section~\ref{sec:experiments}. Related work is presented in Section~\ref{sec:related}. Finally, we conclude the paper 
in Section~\ref{sec:conclusion}.

\vspace{-2mm}
\section{Background} 
\label{sec:background}

The mathematical notations used in this paper are summarised in Table~\ref{tab:notation}.



\begin{table}[th]
\footnotesize
\centering
\caption{Mathematical Notations}
\begin{tabular}{| l | p{6.5cm} |}
\hline 
 Symbol    & Description \\ \hline
$M$ &Number of links in the network\\
$N$ & Number of voxels in the AoI \\ 
$R_i(t)$ & RSS measurement for link $i$ at time $t$ (omni only) \\ 
$\bar{R}_i$ & Mean RSS measurement for link $i$ during calibration (omni only) \\ 
$y_i(t)$ & link quality statistics for link $i$ at time $t$ \\
$y$ & a ${M}\times{1}$ vector of link quality statistics  \\
$x$ & a ${N}\times{1}$ vector of tomographic image \\
$F_i$ & The set of selected \emph{Pattern Pair}s for link $i$ \\ 
$R_{i,j}(t)$ & RSS measurement for link $i$, \emph{Pattern Pair} $j$ at time $t$ (dRTI only) \\
$\bar{R}_{i,j}$ & Mean RSS measurement for link $i$, \emph{Pattern Pair} $j$ during calibration (dRTI only) \\
\hline
\end{tabular}
\label{tab:notation}
\end{table}

\vspace{-2mm}
\subsection{Radio Tomographic Imaging}
\label{subsec:bgRTI}


The aim of RTI is to localise the people and objects (Note: we will also use obstructions to refer to people or objects because this is how RTI ``sees" them.) within an AoI by using sensors placed around the periphery of the AoI. We assume that the AoI is divided into voxels, see Fig.~\ref{fig:example1}. The sensors exchange packets periodically in order to monitor the Received Signal Strength (RSS) of the links over time. In RTI, all links are assumed to be \emph{asymmetric}. 

RTI localises the obstructions in the AoI in two steps. A \emph{tomographic image} is first computed from the RSS measurements, and then a Kalman filter uses the computed image to localise and track the obstructions.  

We use the RTI model introduced in~\cite{WilsonRTI:2010} to compute the tomographic image. This model is also used in many other RTI works~\cite{KaltiokallioBP12,Kaltiokalliograndma2012,maas2013toward,WilsonVRTI:2011, ZhaoKRTI:2013}. Let $R_i(t)$ denote the RSS measurement of the $i$-th link at time $t$. We use $R_i(t)$ to compute some link quality statistics $y_i(t)$ for link $i$. These link statistics $y_i(t)$ will be used to compute the tomographic image later on. We consider the following link quality statistics in this paper. 

\textbf{Mean based RTI (mRTI):}
The \emph{original} mRTI method is proposed in~\cite{WilsonRTI:2010} where RTI is used in an \emph{open} outdoor environment. This is a fairly simple radio environment where radio signal attenuation is due mainly to LOS path loss. The effect of an obstruction in a radio link is to decrease the RSS values of that link. Therefore, one may detect the presence of an obstruction in the link $i$ at time $t$ by testing whether $R_i(t)$ has decreased from a base value. The mRTI method requires a calibration period where obstructions are absent in the AoI. During this period, the sensors exchange packets in order to determine the mean RSS of each link. Let $\bar{R}_i$ be the mean RSS of link $i$ over the calibration period. The original mRTI uses the link statistics $y_i(t) = R_i(t) - \bar{R}_i$ for link $i$. 

We will be concerned with RTI in an indoor environment in this paper. We will see later on that, in the indoor environment, the presence of an obstruction in a radio link can cause the RSS of the link to decrease, increase or stay at the same value. The original mRTI therefore does not work well. In this paper, we use mRTI to mean the \emph{modified} link statistics $y_i(t) = |R_i(t) - \bar{R}_i|$ where absolute value is used because the RSS can increase or decrease. 


\textbf{Variance based RTI (vRTI):}
The vRTI~\cite{WilsonVRTI:2011} method uses the variance of a window of $v$ measurements as the link quality statistics, i.e. $y_i(t) = var(R_i(t), ...,$ $R_i(t - v + 1))$. Since this method does not require a base value, no calibrations are required. 

We collect the $y_i(t)$ for all the $M$ links in the WSN to form the link quality statistics vector $y$. We use $y$ to estimate the tomographic image $x$ which is a $N \times 1$ vector where $N$ is the number of voxels in the AoI. Each element of $x$ corresponds to a voxel. If the $i$-th element of $x$ has a larger value, then the chance of the obstruction \bo{is} in the $i$-th voxel is higher. Fig.~\ref{fig:example3} shows an example tomographic image as an heat map (best view in colour). It can be seen that the true location of the obstruction, which is marked by a cross, is in an ``hot" area. To obtain $x$ from $y$, the RTI model in~\cite{WilsonRTI:2010} assumes that $y$ and $x$ are linearly related: 
\begin{equation}
y = A x + n  \label{eqn:rtimodel}
\end{equation}
where $n$ is a noise vector and $A$ is a $M \times N$ matrix. The $(i,j)$ element of $A$ is given by~\cite{WilsonRTI:2010,WilsonVRTI:2011}: 
\begin{equation}
A_{ij} =  \frac{1}{ \sqrt{d} } \begin{cases}1 & \quad \text{ if } d_{ij}(1) + d_{ij}(2) < d +  \lambda \\0 & \quad \text{ otherwise }  \end{cases}, \label{eqn:rtiweight}
\end{equation}
where $d$ is the distance between the nodes of link $i$, $d_{ij}(1)$ and $d_{ij}(2)$ are the distances from the voxel $j$ to the two nodes of link $i$, and $\lambda$ is a parameter to tune the width of the ellipse. If $A_{ij}$ is non-zero, then it means voxel $j$ is close to the line connecting the nodes of link $i$ and is likely to contribute to the variation that is found in the link quality statistics $y_i(t)$. The non-zero elements of $A_{ij}$ for a given link $i$ form an ellipse, see Fig.~\ref{fig:example1} where the shaded voxels indicates those $A_{ij}$ that are non-zero for the link consisting of the two nodes shown in the figure. 

 


Since the number of voxels $N$ is normally greater than the number of links $M$, Eq.~(\ref{eqn:rtimodel}) is under-determined and has to be solved by using Tikhonov regularization. The estimated tomographic image is then input into a Kalman filter for location estimation and tracking. The details can be found in~\cite{WilsonRTI:2010, WilsonVRTI:2011}.

%

\vspace{-2mm}
\subsection{Electronically Switched Directional Antenna}\label{subsec:esd}

ESD antennas are able to electronically control the direction of the
maximum antenna gain, which is feasible for the resource constrained 
WSNs as demonstrated earlier~\cite{giorgetti07directional, nilsson2009directional}.
Nilsson~\cite{nilsson2009directional} has designed an ESD antenna based on the concept of Electrically-Steerable
Parasitic Array Radiator (ESPAR). 
ESPAR antennas consist of a
central monopole surrounded by a number of parasitic elements.
In the simplest form, the parasitic elements can be 
either grounded or isolated allowing them to 
act as reflectors when grounded and as directors
when isolated. 

In this paper, we use ESPAR antennas, each of which has six parasitic elements designed by 
Nilsson~\cite{nilsson2009directional}. The parasitic elements can be
individually grounded or isolated via software APIs.
When all the parasitic elements
are grounded except one, the direction of the maximum antenna gain is
towards the direction of the isolated element.
We use this configuration for the dRTI experiments. 
When all the parasitic
elements are isolated, we have an omni-directional configuration of the
antenna. We use this configuration for omni RTI and cRTI experiments. 
The antenna gain varies as an offset circle from
approximately 7 dB to -4 dB in the horizontal plane. 

\vspace{-2mm}
\section{Radio Tomographic Imaging with Directional Communications}
\label{sec:methods}

\vspace{-2mm}
\subsection{Superiority of Directional Communications}
\label{subsec:superiority}
An important question concerning the use of directional communications in RTI is how directional antennas can produce better RTI results compared to  omni. Since RTI relies on the change of link quality statistics to infer the presence of an obstruction, we can try to answer this question by comparing the impact of obstructions on the link quality in omni-directional and directional communications. The following discussion assumes power is measured in decibels. 


For omni-direction communications, the received power $P_{rx}(i,t)$ of link $i$ at time $t$ is given by~\cite{WilsonRTI:2010}: 
\begin{equation}
P_{rx}(i,t) = P_{tx}(i,t) - A(i,t) + n(i,t) \label{eqn:rssmodel}
\end{equation}
where 
\begin{equation}
A(i,t) = L(i,t) + S(i,t) - F(i,t)  \label{eqn:rssPL}
\end{equation}
where $P_{tx}(i,t)$ is the transmitter transmission power and $n(i,t)$ is the measurement noise. The term $A(i,t)$ is the total attenuation consisting of three parts: path loss $L(i,t)$, shadowing loss $S(i,t)$ and fading loss $F(i,t)$. The RSS measurement $R_i(t)$ for link $i$ is a quantised version of the received power $P_{rx}(i,t)$. In an open outdoor environment, an obstruction causes large shadowing loss and results in a drop in the link RSS. This can be detected by the original mRTI as discussed in Section~\ref{sec:background}. However, the indoor NLOS environment is \emph{drastically} different. Multi-path fading in an indoor environment can cause both constructive and destructive interference. This means that, when an obstruction blocks a radio link, the RSS of that link can \emph{decrease}, \emph{increase}, or even \emph{remain unchanged}. In this situation, vRTI, which uses the variance of RSS as the link quality statistics, performs better than the original mRTI~\cite{WilsonVRTI:2011}.


 \begin{figure*}[htb]
    \subfigure[RTI with omni-directional antennas. ]{
      \includegraphics[scale = 0.42]{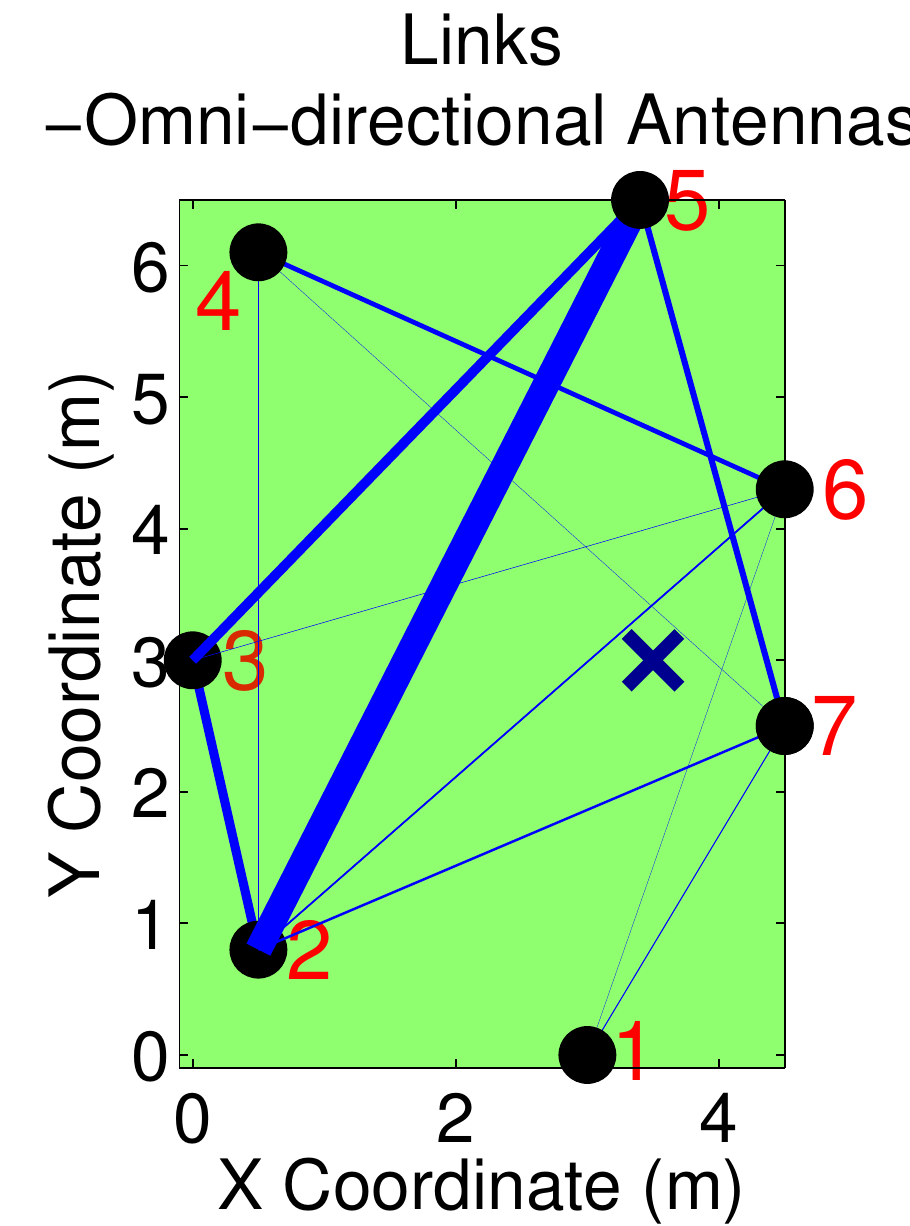}
    \label{fig:rti_omni}
    }
    \subfigure[The produced image from (a) (Picture view best in colour). ]{
     \includegraphics[scale = 0.42]{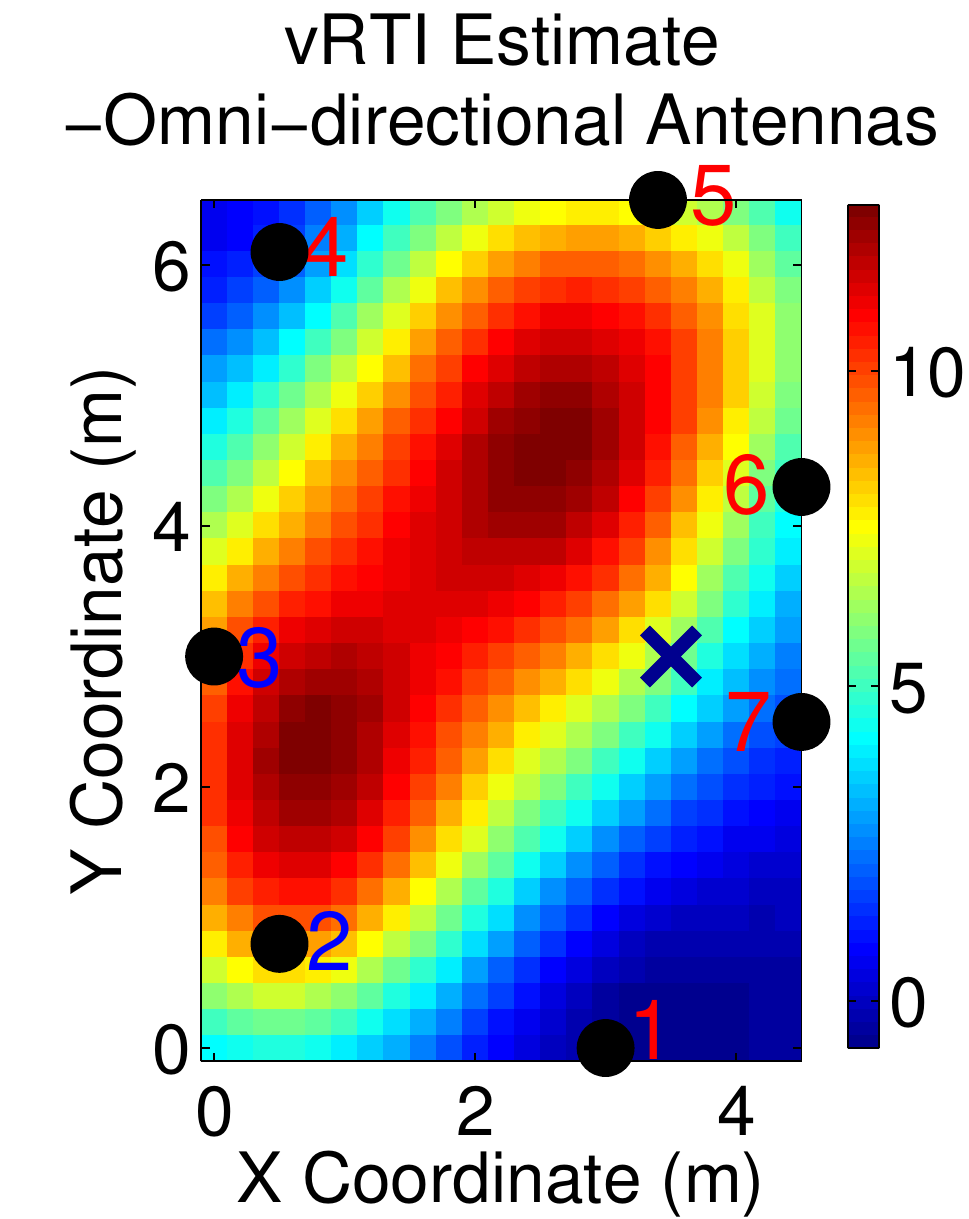}
    \label{fig:image_omni}
    }
   \subfigure[RTI with \emph{directional} antennas.]{
       \includegraphics[scale = 0.42]{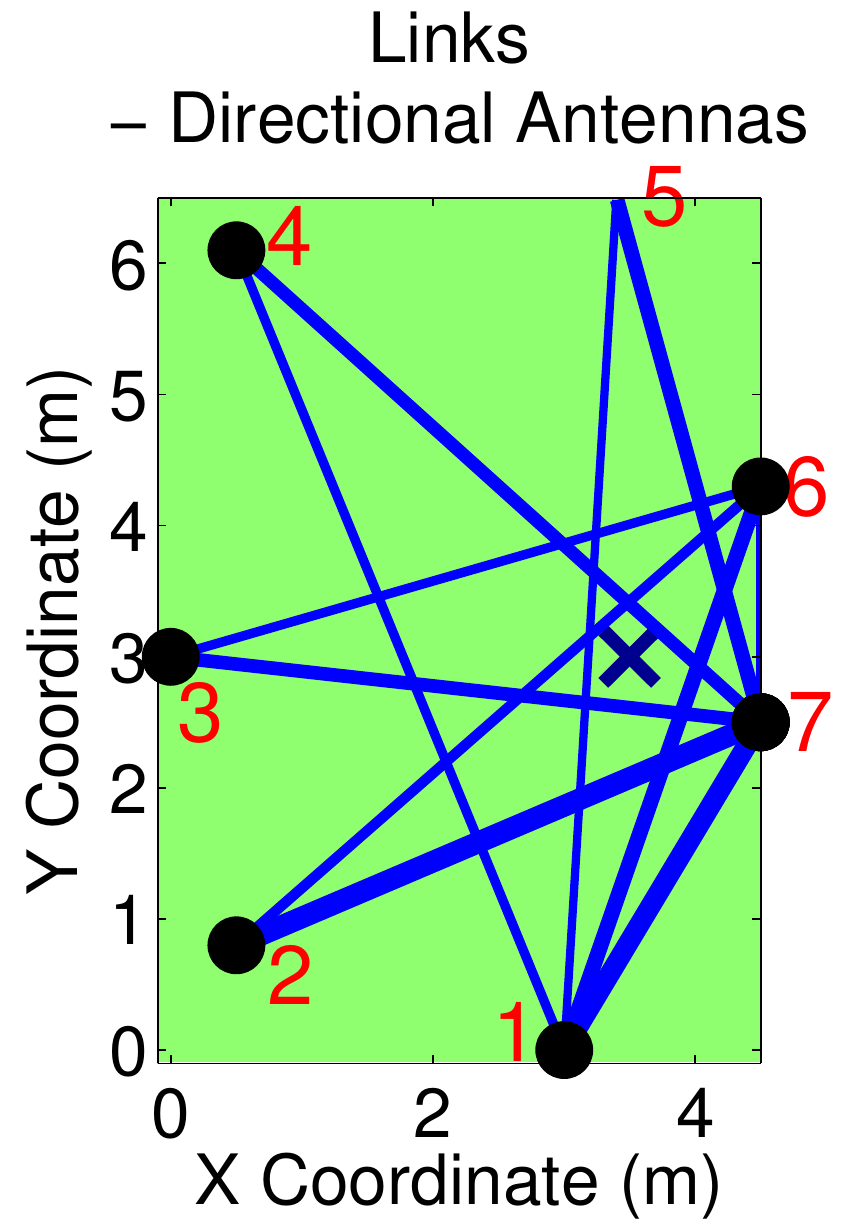}
    \label{fig:rti_dir}
    }
   \subfigure[The produced image from (c) (Picture view best in colour).]{
     \includegraphics[scale = 0.42]{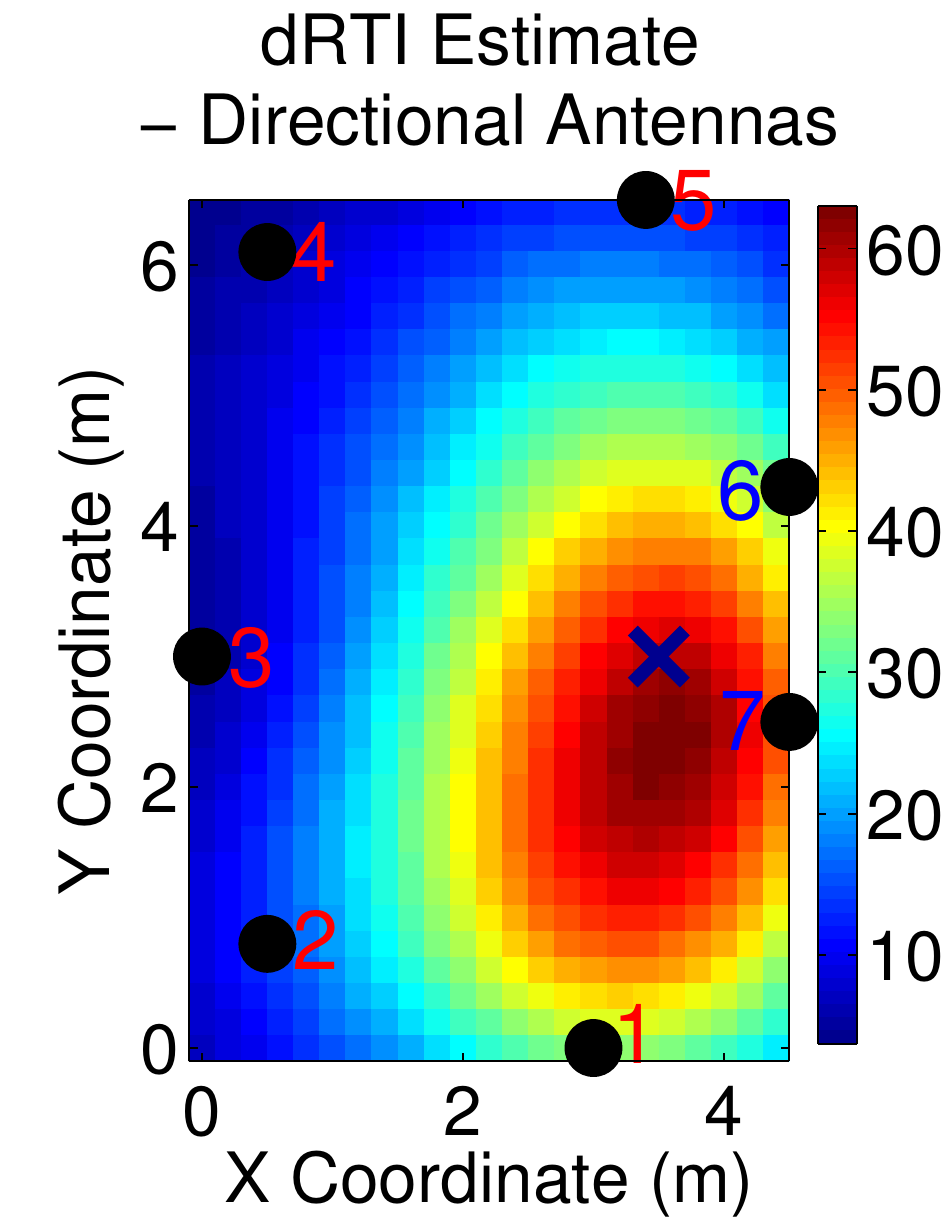}
    \label{fig:image_dir}
    }
    \caption{Omni-directional RTI vs dRTI. The thicknesses of lines in (a) and (c) are proportional to the vRTI link quality statistics. The crosses show the true location of the obstruction. Fig.~(b) shows that the RTI image with a large localisation error because of a deep fade between Nodes 2 and 5 in (a). However, this fade has little impact on dRTI estimation shown in (d). }
    \label{fig:links_dir_vs_omni}
\end{figure*} 

 

For directional communications, the received power $P_{rx}(i,t)$ between a pair of transmission and received antennas (here the index $i$ refers an antenna pair) depends on their relative orientation specified by zenith $\phi$ and azimuth $\theta$. \cite{AndersonDirMol2009} proposes the model: 
 \begin{equation}
P_{rx}(i,t)  \propto f(\phi,\theta)  \times x(\phi,\theta)
\label{eqn:rssdir}
\end{equation}
where $f(\phi,\theta)$ is a direction-specific antenna gain function and $x(\phi,\theta)$ is an environment-specific offset function. The directional-specific gain is often touted as the advantage of using directional antennas for communication because the transmitter and receiver can focus their respective sending and receive antenna gains to boost the signal-to-noise ratio of the communication~\cite{ramanathan2001performance, winters1998smart}. This generally holds for outdoor LOS communication. However, the question of whether directional antennas offer any advantages in an indoor environment, where NLOS and multi-path effects dominate, has only been studied recently. These studies are all empirical because it is difficult to model the environment-specific offset function $x(\phi,\theta)$ analytically. 

In the study by Lakshmanan et al.~\cite{Lakshmanan2010}, the authors place a pair of wireless devices, each equipped with a MIMO antenna array, in an indoor environment. One device acts as the transmitter and the other acts as a receiver. The channel gains from the different antennas on the transmitter device to those on the receiver device are measured. The measurements show that these channel gains vary greatly. This means a slight change in antenna location can cause big changes in channel gain. In another study, Sani et al.~\cite{AmiriSani:2010:DAD:1859995.1860021} measure the channel gain in many different orientations by changing the sending and receive directions of the directional antennas. They find that directional antennas outperform omni over a large range of orientations but are significantly worse over a small range of orientations in an indoor environment. 

What do all these results mean for RTI? First, studies on the impact of obstructions on link quality of directional communication in an indoor environment do not exist. Second, radio propagation in an indoor environment is full of surprises and the only way to find out the truth is through real experiments. We therefore conduct our own experiments to determine, whether and to what extent, dRTI can improve the localisation accuracy. These experiments and their results will be described in this section and the next. 

In order to motivate future sections,
Fig.~\ref{fig:links_dir_vs_omni} compares the localisation error between omni-direction RTI and dRTI. These experiments are conducted within 10 ms of each other so their channel conditions can be assumed to be \emph{comparable}. The RTI image for omni in Fig.~\ref{fig:image_omni} shows a larger error compared to that of dRTI in Fig.~\ref{fig:image_dir}. In order to explain this difference, we look at the vRTI link quality statistics, which are indicated by the thickness of the lines connecting the node pairs, in Fig.~\ref{fig:rti_omni} and Fig.~\ref{fig:rti_dir} for omni and directional. 
Fig.~\ref{fig:rti_omni} shows that a deep fade between Nodes 2 and 5 has led omni-vRTI to believe that an obstruction is likely to be present between these two nodes, as shown by the high value of tomographic image for the voxels around the line joining these nodes. However, this deep fade had little impact on dRTI. 
Our hypothesis is that \emph{directional communication is sensitive to obstruction over a large range of orientations. This results in both larger changes in RSS and larger RSS variance. Moreover, these changes are less affected by fading.} 
We will provide empirical evidence to support this hypothesis next.  
 



\vspace{-2mm}
\subsection{The Impact of Directional Communications on RSS Changes and Variance}
\label{subsec:variance}
This section presents empirical evidence that directional antenna links, when obstructed, produce larger changes in RSS and larger RSS variance in most antenna directions.  

\vspace{-2mm}
\subsubsection{Experimental Setup}
\label{subsubsec:setup}
We employ two TelosB nodes running the Contiki operating system. Each node has an ESD antenna with 6 parasitic elements equally spaced at 60$^{\circ}$ apart. The nodes can \emph{dynamically} change the sending direction to one of the six directions or send omni-directionally by setting a digital output pin. It takes less than 1ms to change the sending direction. The RSS Indicator (RSSI) values are read from the TI CC2420 transceivers and converted to RSS values in dBm according to the transceiver datasheet. 

\begin{figure}[htb]
   \centering
    \includegraphics[scale = 0.45]{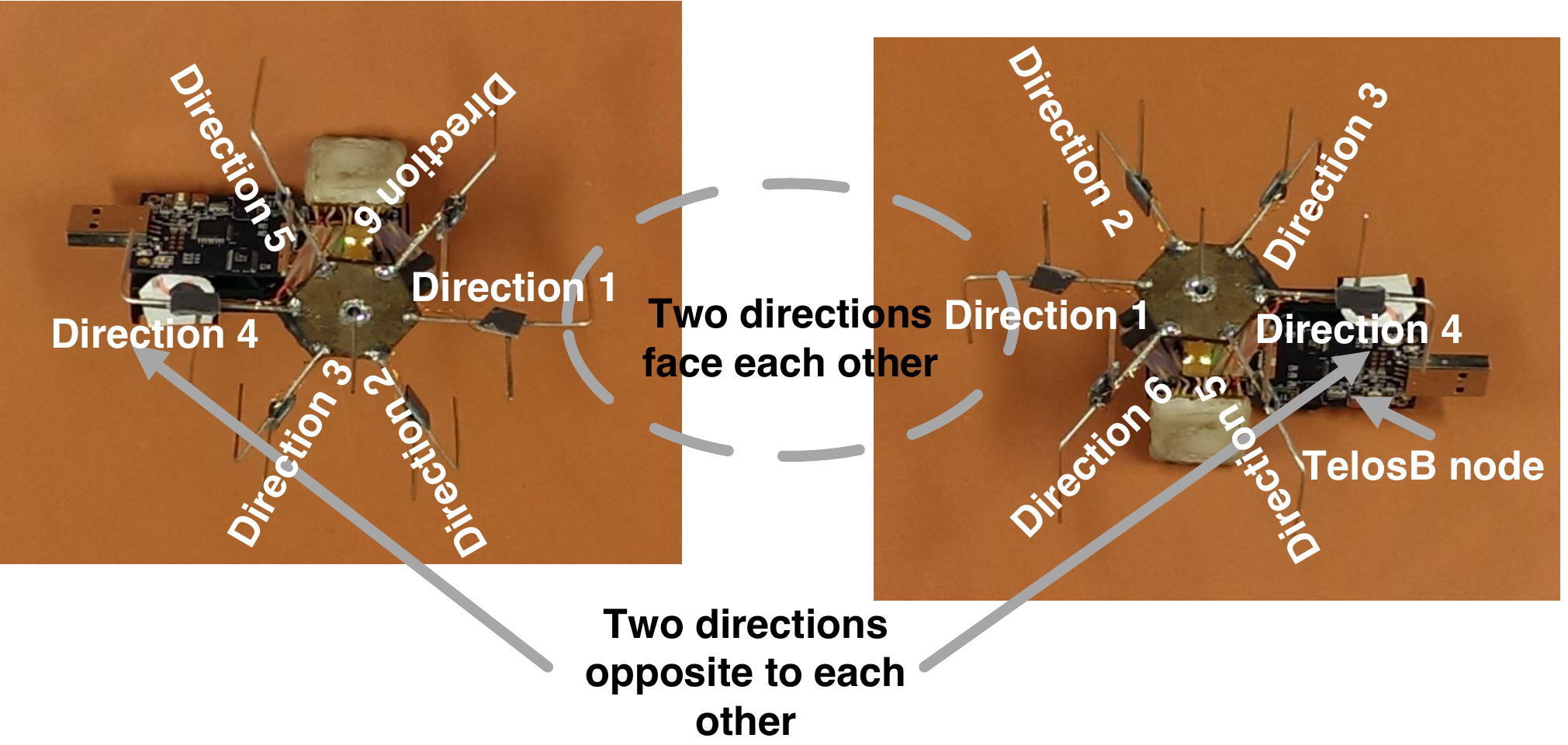}
    \caption{The physical ESD antenna directions of two nodes}\label{fig:antennafacing}
\end{figure}

We conduct experiments in both indoor LOS and NLOS (``through-wall'') environments in our lab. We only report the results of the NLOS experiment here because we observe similar results for the LOS environment. For the NLOS environment, we place the two nodes, approximately 3 meters apart, in two different rooms. The wall between the rooms is made of glass and wood. We label the antenna directions with a number from 1 to 6. The \emph{Direction 1}s of the two nodes face each other, see Fig.~\ref{fig:antennafacing}. A person walks through the link of the two nodes \emph{multiple times} to block it. 

The transmitter sends small packets to the receiver in each of the six directions, as well as omni-direction as fast as possible, and the receiver receives the packets from all six directions and omni-direction. The transmission node ID, direction, and the sequence number are included in the packets. When the receiver receives a packet, it appends the receive direction and RSSI value into the packet as well. Therefore, we can figure out the transmit direction, receive direction and RSS value of each received packet. 
For a radio link from a transmitter to a receiver, there are 36 possible \emph{transmit-receive antenna direction pairs} and we will refer to them as  \emph{Pattern Pairs}. Furthermore, there are two links (uplink and downlink) between two nodes.

\begin{figure*}[!ht]
  \centering
    \subfigure[RSS measurement change]{
        \includegraphics[scale = .20]{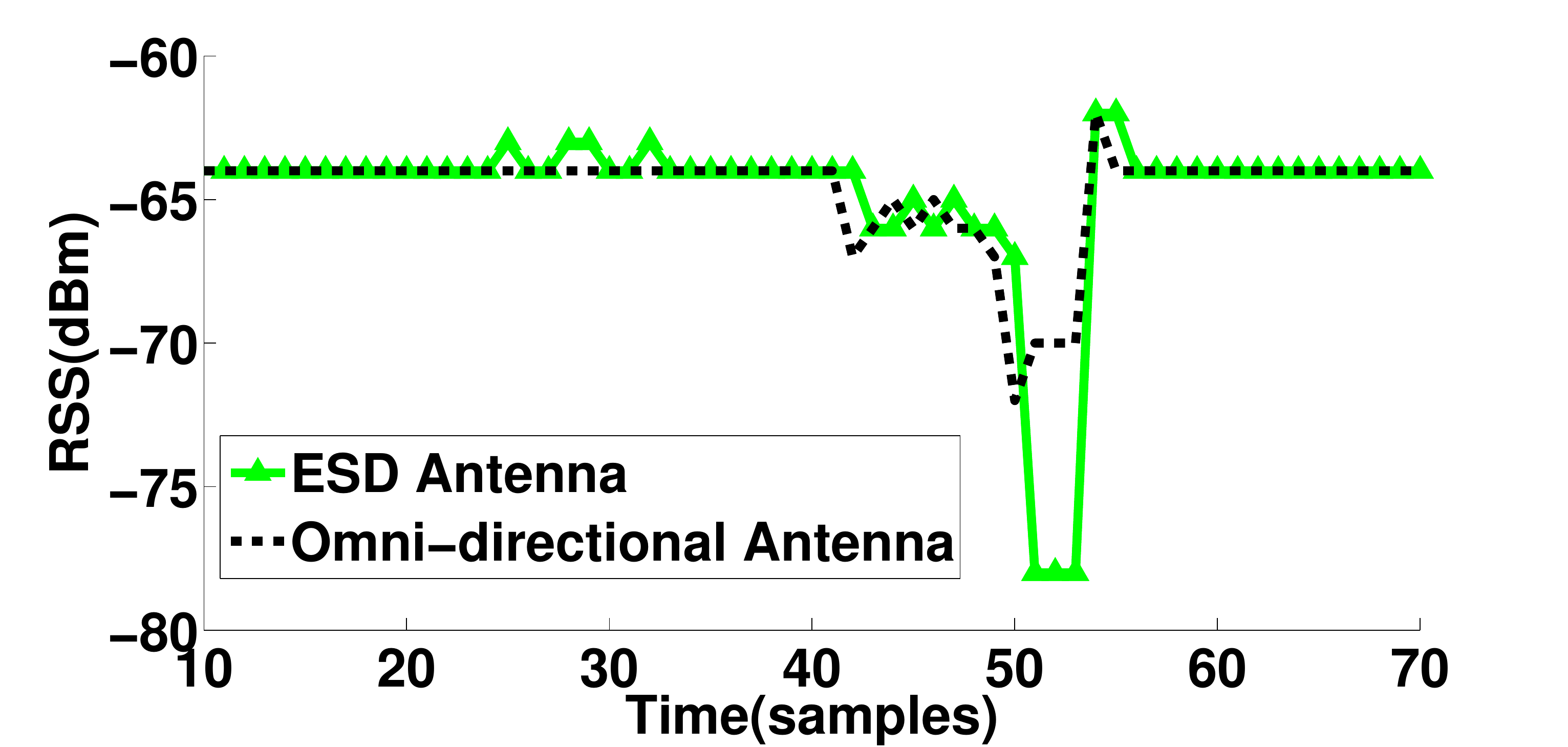}
        \label{fig:largervarRSS}
    }
     \subfigure[mRTI]{
    \includegraphics[scale = 0.25]{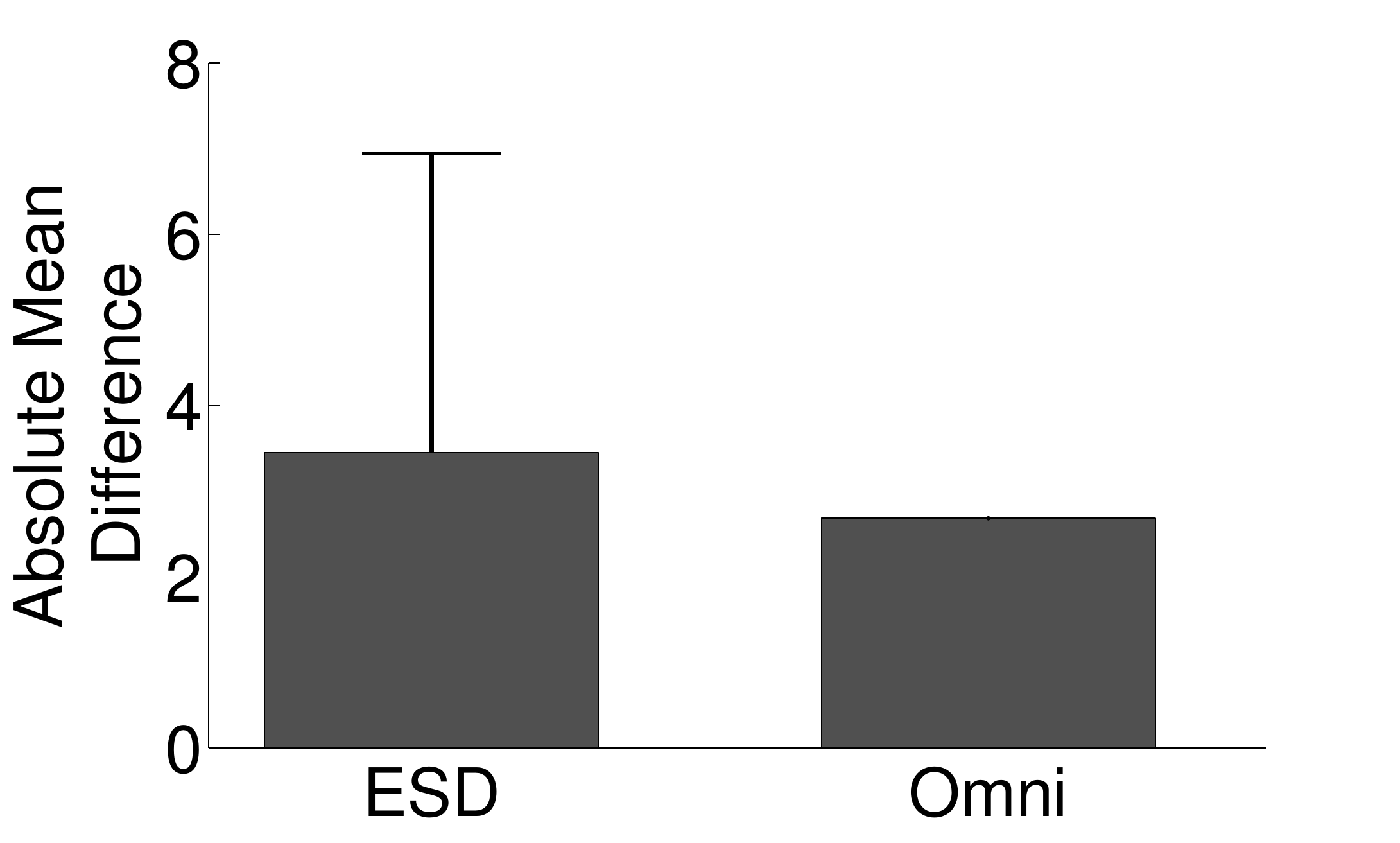}  
    \label{fig:largermeansta}
    }
    \subfigure[vRTI]{
    \includegraphics[scale = 0.25]{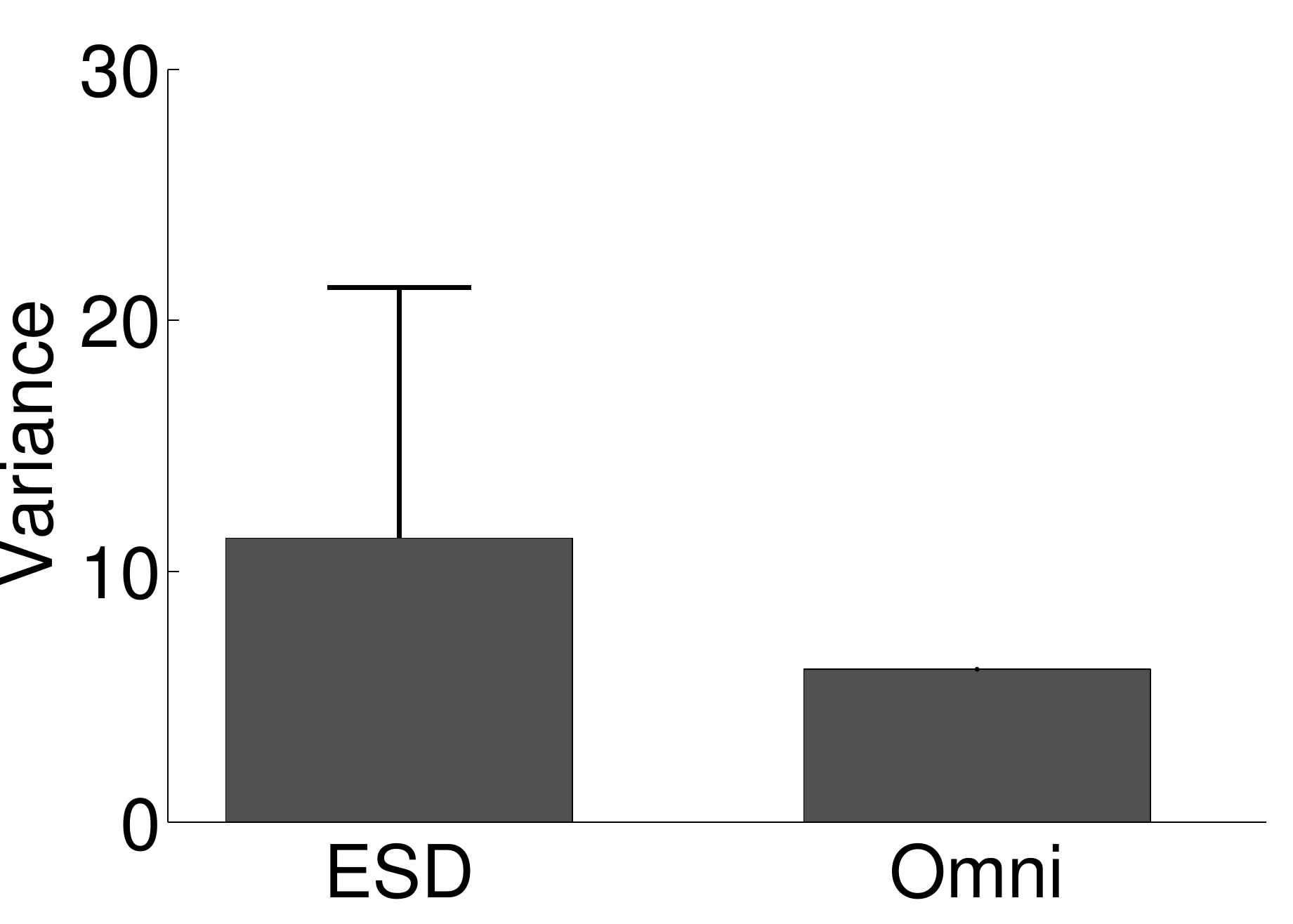}  
    \label{fig:largervarsta}
    }
    \caption{ (a) RSS measurement of a selected \emph{Pattern Pair} and the omni-directional antenna. (b) The average mRTI over 36 \emph{Pattern Pairs} versus mRTI of omni. (c) The average vRTI over 36  \emph{Pattern Pairs} versus vRTI of omni. The error bars indicate one standard deviation.}\label{fig:largervariance}
\end{figure*} 

\vspace{-2mm}
\subsubsection{The Impact to RSS Changes and Variance}
Fig.~\ref{fig:largervarRSS} shows the RSS measurements of one of the \emph{Pattern Pairs} and omni when a person attenuates the radio signal between time 50 and 55. It is clear that the \emph{Pattern Pair} has a significantly bigger change in RSS measurement than omni. We compute the mRTI and vRTI of the RSS measurements for the 36 \emph{Pattern Pairs} and omni. (Note that mRTI and vRTI in Section \ref{subsec:bgRTI} are defined for a link. We simply consider a \emph{Pattern Pair} as a link in the calculation.) We compare the average mRTI and vRTI over the 36 \emph{Pattern Pairs} against that of omni in Fig.~\ref{fig:largermeansta} and \ref{fig:largervarsta}. These figures show that both the change in RSS and variance of RSS are bigger for directional antennas. In particular, the average RSS variance over all 36 \emph{Pattern Pairs} is almost twice that of omni. This experiment demonstrates that \emph{directional communications are indeed more sensitive to the obstruction, and produce larger changes in RSS and larger changes in RSS variance}.
Fig.~\ref{fig:largermeansta} and \ref{fig:largervarsta} also show the standard deviation of the mRTI and vRTI over the 36 \emph{Pattern Pairs}. The standard deviation is of almost the same magnitude as the average. We will discuss the variation of mRTI and vRTI among the \emph{Pattern Pairs} later.

\vspace{-2mm}
\subsubsection{Radio Link Attenuation False Negatives and False Positives} \label{sec:hypoFNFP}
  
We say that a radio link attenuation \emph{false negative}\bo{\emph{(FN)}} has occurred if the RSS measurements remain stable (i.e., with little change and small variance) in the presence of an obstruction in the LOS path (e.g., within the shaded area in Fig.~\ref{fig:example1}). Similarly, we say that a radio link attenuation \emph{false positive}\bo{\emph{(FP)}} has occurred if the link quality RSS measurements change significantly when no obstructions are present in the LOS path. Both \emph{false negatives} and \emph{false positives} frequently occur in RTI systems, especially in the NLOS cluttered environments. These errors can significantly impact on the accuracy of RTI localisation. 

\begin{figure}[htb]
    \centering
        \subfigure[example of false positive]{
        \includegraphics[scale = .25]{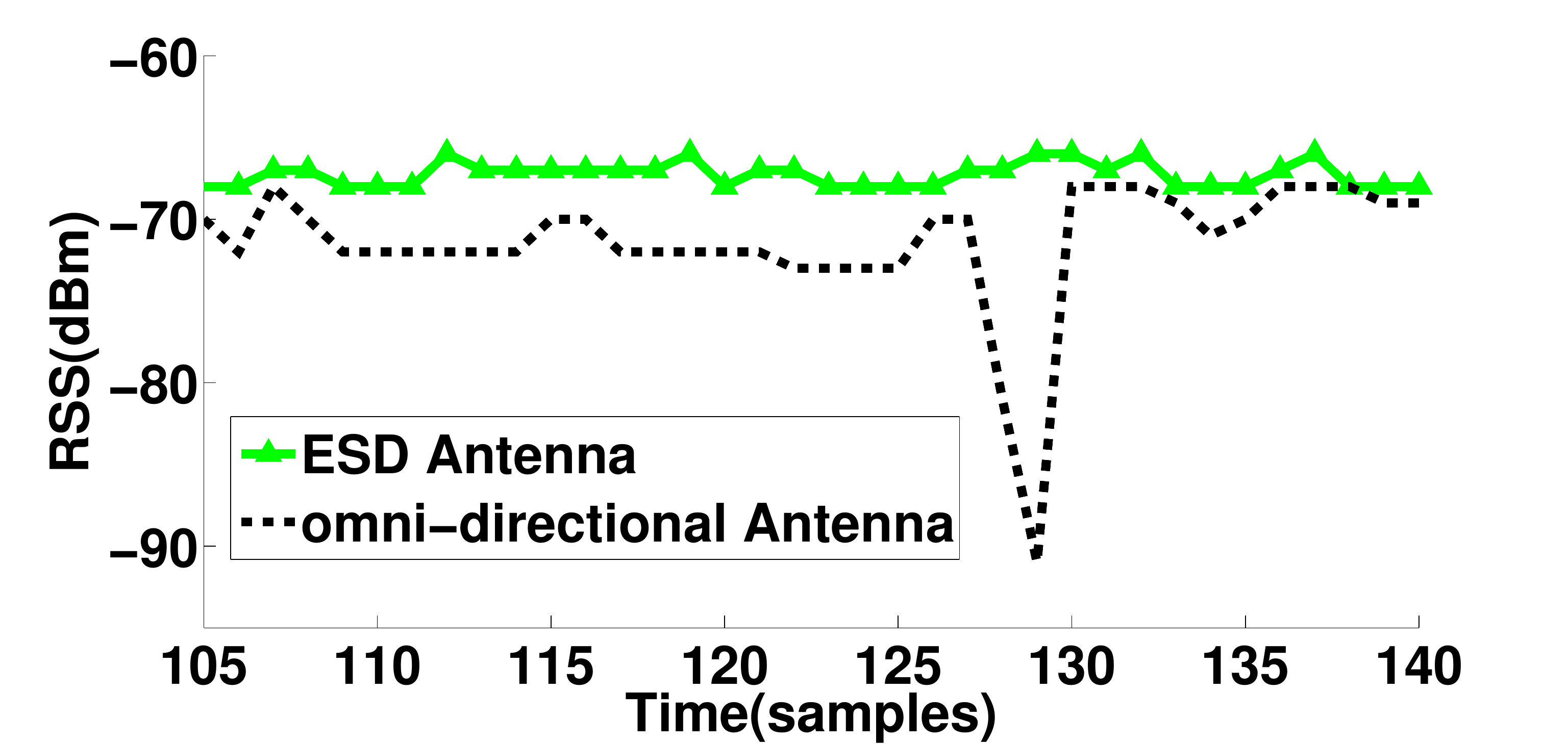}
        \label{fig:FP}
    }
    \subfigure[example of false negative]{
    \includegraphics[scale = .25]{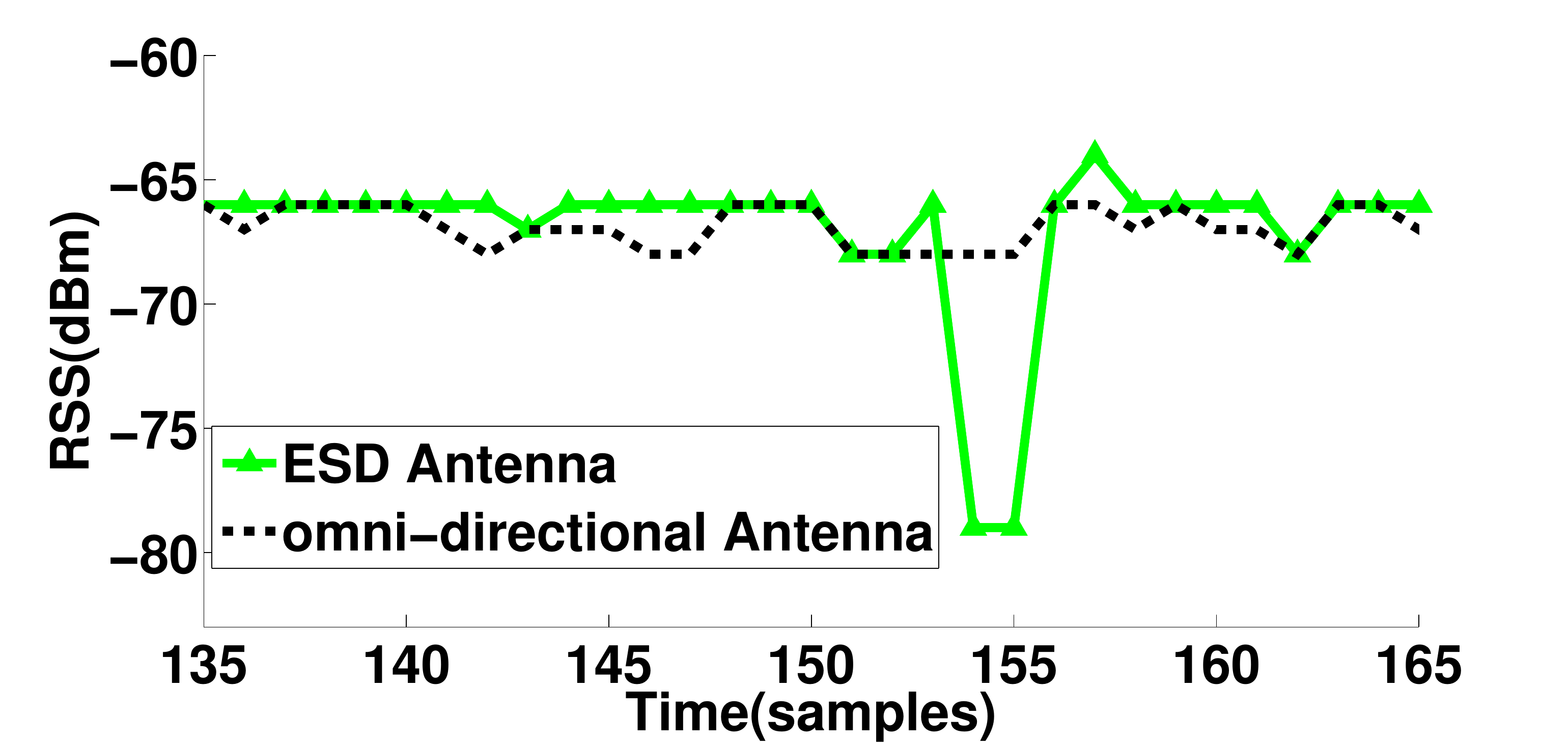}    
    \label{fig:FN}
    }
    \caption{Examples of false positive and false negative. (a) Omni direction communication produces a \emph{false positive} without the presence of obstructions; (b) Omni direction communication produces a \emph{false negative} in the presence of an obstruction.}\label{fig:FPFN}
\end{figure}

A benefit of directional antennas is that they can reduce the instances of false positives and negatives. Fig.~\ref{fig:FPFN} shows examples of the RSS measurement from experiments when omni-direction communications produce \emph{false positives} (see Fig.~\ref{fig:FP}) and \emph{false negatives} (see Fig.~\ref{fig:FN}), but not for \emph{some} directional \emph{Pattern Pairs}. 
As a result, direction communications can indeed improve the accuracy of RTI systems. The poor localisation accuracy of Fig.~\ref{fig:image_omni} for omni directional antenna is a result of a \emph{false positive} in the link quality between Nodes 2 and 5, and \emph{false negative} in the link quality near Nodes 6 and 7. We have observed many instances of such behaviour in other experiments. 
\bo{See Section~\ref{subsec:comprti} for the details about the statistics in the experiments.}
 
\vspace{-2mm}
\subsection{Directional \emph{Pattern Pair} Selection}
\label{subsec:antennachoose}

The number of \emph{Pattern Pairs} between two nodes can be large. Each of the nodes that we use in this paper has an ESD antenna with 6 parasitic elements. The maximum number of \emph{Pattern Pairs} between 2 nodes is $36 \times 2 = 72$ where the $2$ comes from the uplink and downlink. This means  there are significant transmission overhead to make RSS measurements from all \emph{Pattern Pairs}. This may make the dRTI system difficult to scale to larger sizes.

\begin{figure}[htb]
    \centering
       \subfigure[mRTI]{
    \includegraphics[scale = 0.2]{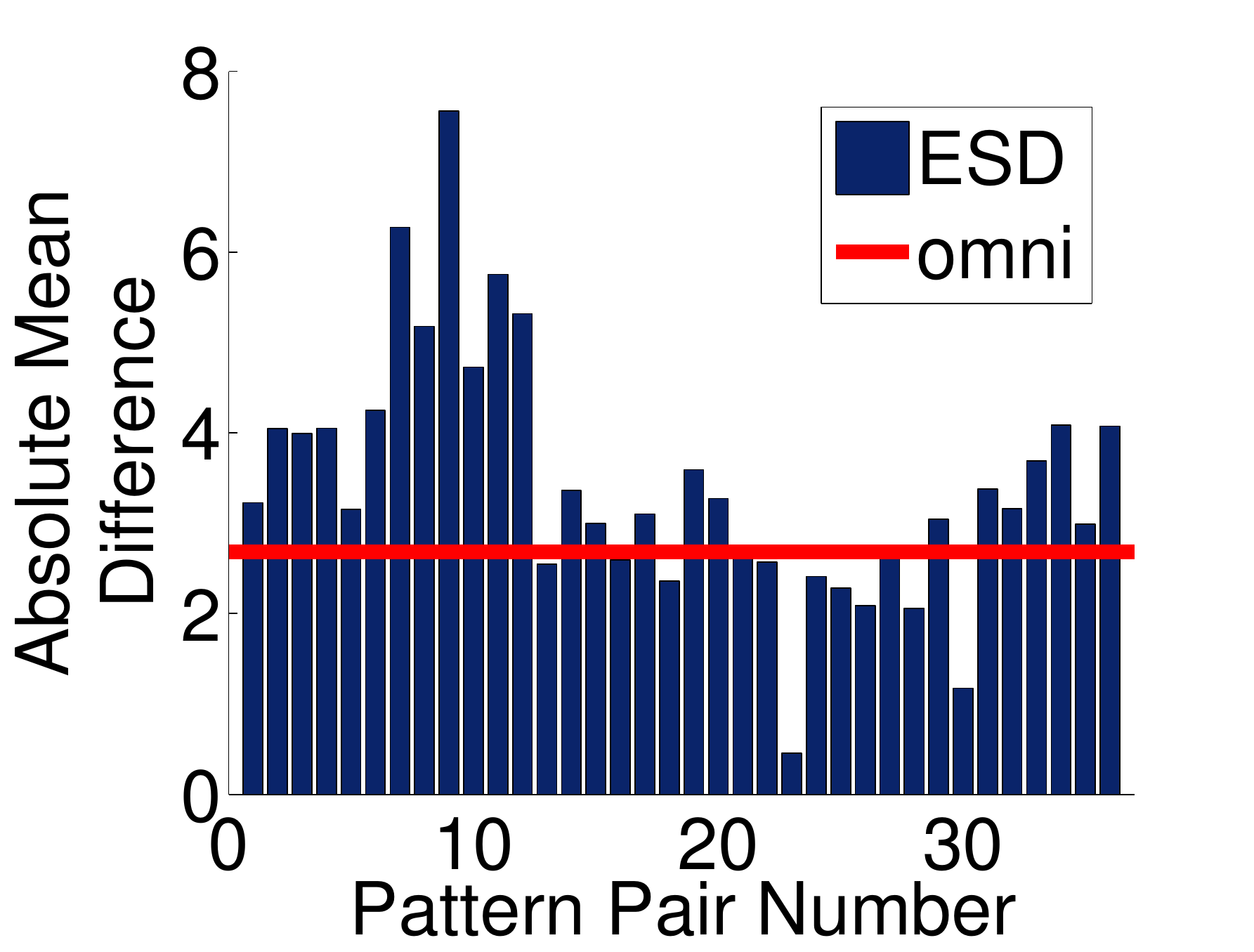}  
    \label{fig:patternpairMean}
    }
    \subfigure[vRTI]{
    \includegraphics[scale = 0.2]{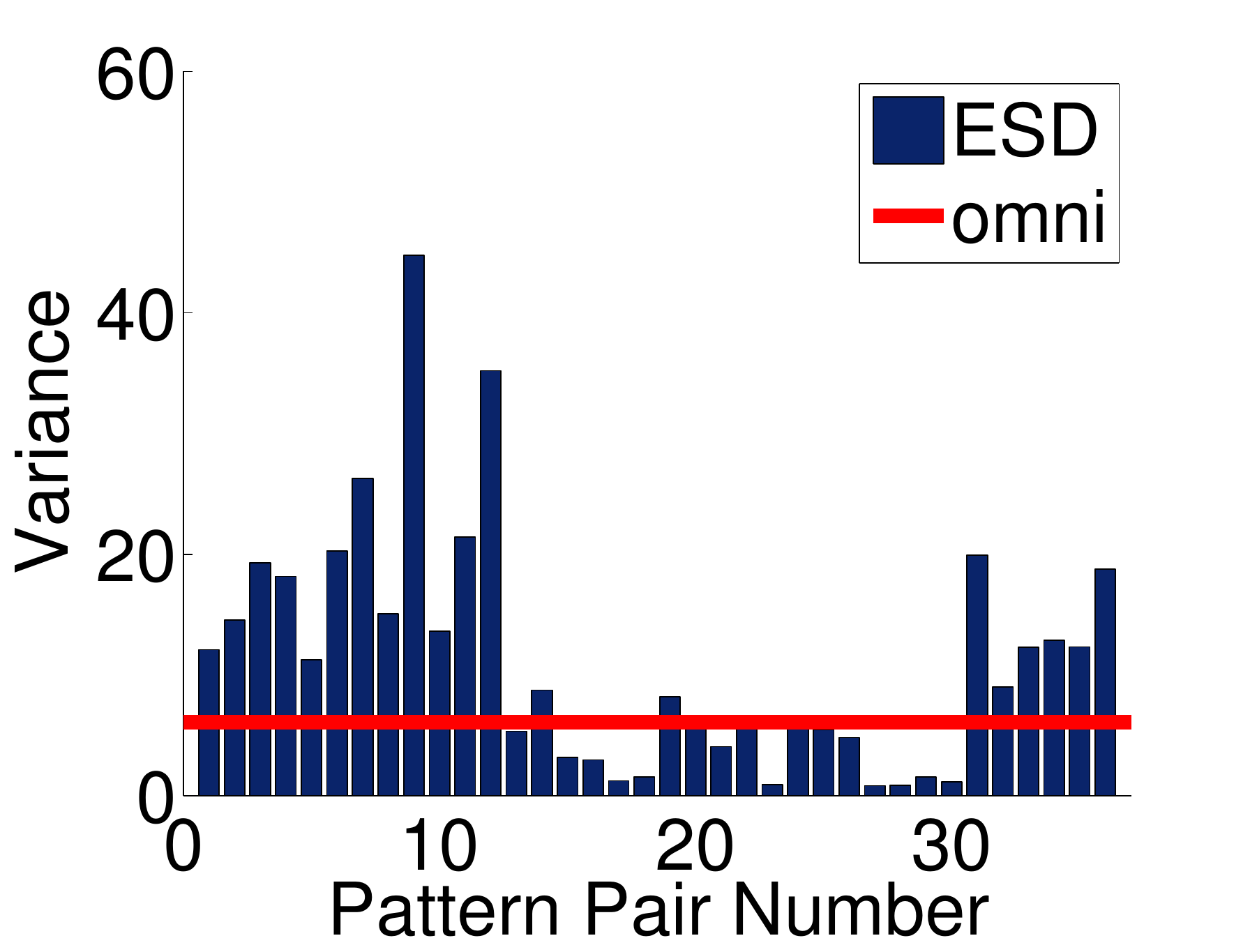}  
    \label{fig:patternpairVar}
    }
    \caption{The mRTI (a) and vRTI (b) link quality statistics for the 36 \emph{Pattern Pairs}. The horizontal line shows the mRTI (a) and vRTI (b) for omni. 
}
    \label{fig:patternpairBoth}
\end{figure}

Furthermore, intuitively, one may think that the \emph{Pattern Pair} with antennas directly pointing to each other (e.g. \emph{Direction 1}s in Fig.~\ref{fig:antennafacing}) should be the ``best'' choice. \emph{Is that true?} Let us look at the mRTI and vRTI for all 36  \emph{Pattern Pairs} in Fig.~\ref{fig:patternpairMean} and Fig.~\ref{fig:patternpairVar} respectively. The answer is that the intuition is \emph{not true}, similar to the earlier finding in wireless mesh networks~\cite{Subramanian:2009:ECS:1530748.1530784}. 
For example, for vRTI in Fig.~\ref{fig:patternpairVar}, the \emph{Pattern Pair} with antennas directly pointing to each other is the first bar from the left which does not have the maximum RSS variance when the link is blocked by an obstruction. However, if both nodes choose Direction 6 (see Fig.~\ref{fig:antennafacing}), this gives the first bar from the right in Fig.~\ref{fig:patternpairVar} which has a higher vRTI. The maximum variance is produced by the \emph{Pattern Pair} between Direction 2 of the receiver and Direction 3 of the transmitter, which is the ninth bar from left in Fig.~\ref{fig:patternpairVar}. 

Similar to the earlier discovery in~\cite{AmiriSani:2010:DAD:1859995.1860021,Lakshmanan2010}, Fig.~\ref{fig:patternpairMean} and \ref{fig:patternpairVar} also show: (1) The link qualities vary significantly between the \emph{Pattern Pairs}; (2) Some \emph{Pattern Pairs} have bigger change/variance than omni but some do not. Therefore, we need to investigate intelligent methods to select the best directional 
\emph{Pattern Pair}(s) to increase the localisation accuracy and to decrease the transmission overhead
 of dRTI systems.
To this end, we introduce three selection methods to choose directional link \emph{Pattern Pair}(s).



%

\vspace{-2mm}
\subsubsection{Location Method}
This method chooses the best \emph{Pattern Pairs} based on the physical location and orientation of the antennas. Intuitively, the \emph{pattern pair} whose antennas are pointing to each other (resp. pointing in the opposite directions) should have the strongest (weakest) signal strength at the receivers, and also have higher (lower) probability to show significant RSS variances when the link between the transmitter and the receiver nodes is blocked by an obstruction. 
%
%
Though our earlier discussion shows that this intuition does not produce the best \emph{Pattern Pair}. Nevertheless, we find from Fig.~\ref{fig:patternpairMean} and \ref{fig:patternpairVar} that those \emph{Pattern Pairs} that are roughly facing each other do generally give a significantly higher link quality statistics (mRTI or vRTI) than omni directional antennas. Furthermore, this selection method \emph{does not require an initial calibration phrase} as the other methods we will introduce, which is important for some applications such as emergency response~\cite{maas2013toward}.
 
\begin{figure}[ht]
    \centering
    \includegraphics[scale = 0.25]{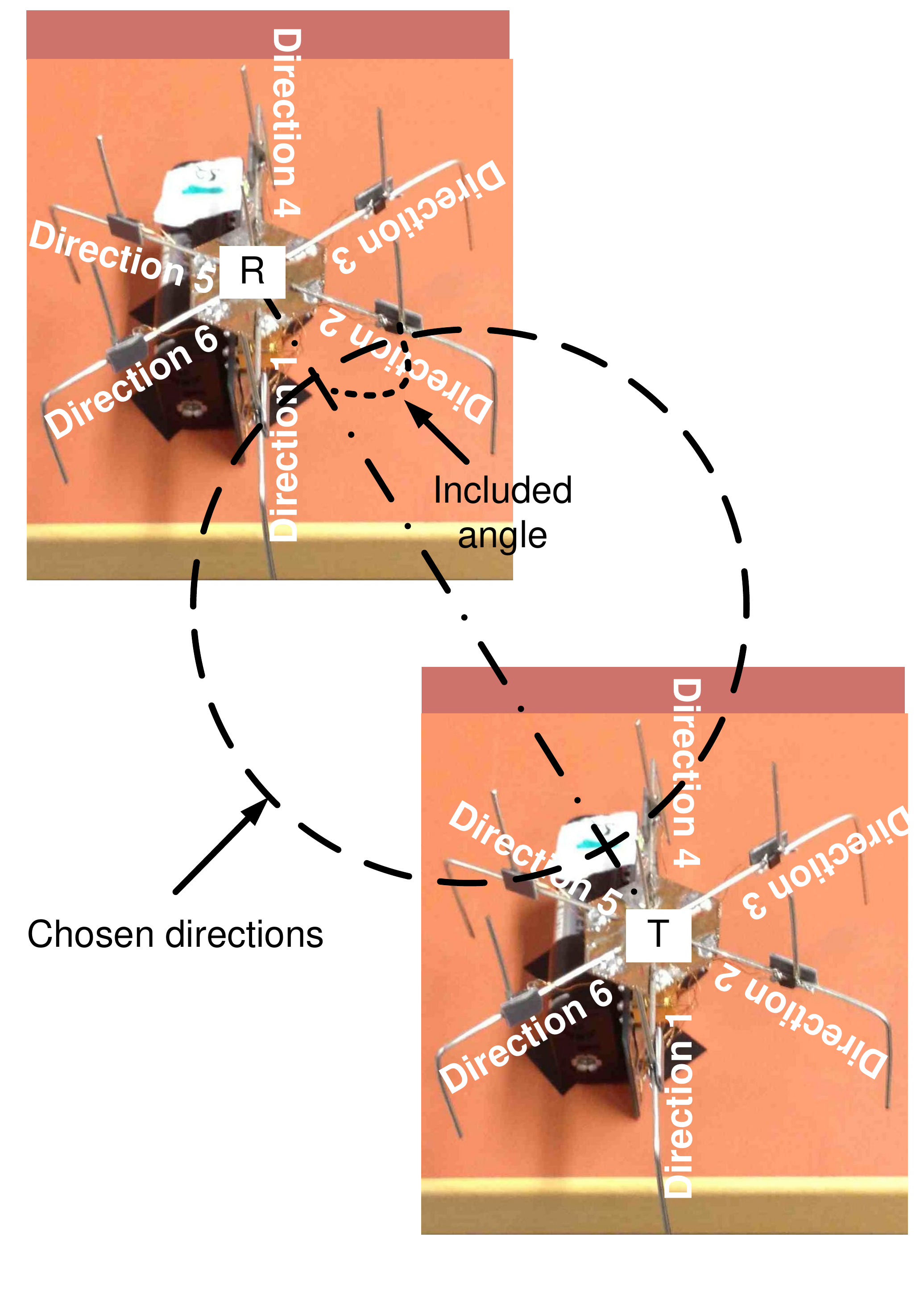}
    \vspace{-5mm}
    \caption{The figure shows the 2 directions of receiver and transmitter chosen by the Location method.}
    \label{fig:patternchoice}
\end{figure}

For this method, we record the orientation of antennas during deployment stage. Then, according to the physical orientation of the antennas, we select $n_{transmitter}$ directions for the transmitting node $N_1$ and $n_{receiver}$ directions for the receiving node $N_2$. Let $N_1 N_2$ denote the line connecting nodes $N_1$ and $N_2$. For the transmitter node $N_1$, we compute the angles (magnitude only) between the antennas on node $N_1$ and the line $N_1 N_2$, and select the directions corresponding to the $n_{transmitter}$ smallest angles. Similar method applies to choosing $n_{receiver}$ directions on $N_2$. In practice, we choose $n_{transmitter} = n_{receiver}$ because each node has to be transmitter and receiver. This means the number of \emph{Pattern Pairs} selected is always a square number. Fig.~\ref{fig:patternchoice} shows an example with $n_{transmitter} = n_{receiver} = 2$.  

%
%

 \vspace{-2mm}

\vspace{-2mm}
\subsubsection{Fade Level Method}
\label{subsubsec:fade_level_method}
Wilson and Patwari define fade level in a radio link as a continuous function between the two extremes of \emph{deep fade} and \emph{anti-fade} in~\cite{wilson2012fade}. A link in deep fade will experience high variance as a person moves in a \emph{wide} area,
which may produce \emph{false positives} when the person is outside the LOS path of a \emph{Pattern Pair}, e.g., outside the shadowed area in Fig.~\ref{fig:example1}. On the other hand, an anti fade \emph{Pattern Pair} is affected by constructive
multi path interference.  The link quality varies significantly less, but the radio signal attenuates significantly when a
person is present in the LOS of the \emph{Pattern Pair}. Therefore, deep fade \emph{Pattern Pair} is less informative for dRTI,
and we would like to select the \emph{Pattern Pairs} that have minimum deep fade level as in~\cite{KaltiokallioBP12}.

The Fade Level method collects RSS measurements of all the \emph{Pattern Pairs} with an \emph{empty} AoI during the calibration period $t \in [t_1, t_2]$. 
During the calibration, 
the transmitter node sends a number of (e.g., 6) packets sequentially in each direction so that receiver nodes can at least receive one packet in each antenna direction. 
After a fixed number of rounds (e.g.,  50), the Fade Level method chooses the \emph{Pattern Pairs} with small fade levels. 
Specifically, for a \emph{Pattern Pair} between direction $i$ of a transmitter $N_{1}$ and 
direction $j$ of a receiver $N_{2}$,
it computes the \emph{normalised} RSS measurements $h_{i,j}$ during calibration period ($t \in$ $[t_1, t_2]$) as 
\begin{equation}
h_{i,j} =  \sum_{t=t_1}^{t_2}{r_{i,j}(t)} 
\label{equ:fade_level}
\end{equation}
where 
\begin{equation}
r_{i,j}(t) =  P_{rx}(i, j, t) - P_{tx}(i, t) 
\label{equ:normalised_rss}
\end{equation}
where $P_{tx}(i, t)$ is the transmission power of $N_1$ in direction $i$ at time $t$, and $P_{rx}(i, j, t)$ is the RSS measurement 
of receiver $N_2$ at direction $j$ for the packets transmitted from $N_1$ in direction $i$ at time $t$. 

As in~\cite{KaltiokallioBP12}, we 
use $h_{i,j}$ as a measure of the fade level, and if $h_{i,j}$ is smaller, the \emph{Pattern Pair}
is in a deeper fade. Therefore, the 
Fade Level method selects the top $k$ \emph{Pattern Pairs} with maximum 
\emph{normalised} RSS measurements ($h_{i,j}$) from the 36 \emph{Pattern Pairs} between $N_1$ to $N_2$.
 

\vspace{-2mm}
\subsubsection{Packet Reception Rate (PRR) Method}
Intuitively, Packet Reception Rate (PRR) is a also good proxy for link quality because good quality links
tend to have higher PRR. The PRR method also has the same calibrating period where the PRR for all \emph{Pattern Pairs} are measured. The PRR method selects the $k$ \emph{Pattern Pairs} between transmitter node $N_1$ and receiver node $N_2$ that have the highest PRR during the calibration period.

\vspace{-2mm}
\subsection{Link Quality Statistics for dRTI} \label{subsec:observation}

In Section \ref{subsec:bgRTI} we present the mean and variance based link quality statistics for omni RTI. We adopt similar statistics for dRTI but they have to be modified to take into account the \emph{Pattern Pairs}. Let $F_i$ denote the set of selected \emph{Pattern Pairs} for link $i$. Let $R_{i,j}(t)$ denote the RSS measurement at time $t$ for the $j$-th \emph{Pattern Pair} of link $i$. The link quality statistics for dRTI are:

\begin{itemize}
\item  \textbf{RSS mean}
Let $\bar{R}_{i,j}$ be the mean RSS over the calibration period for the $j$-th \emph{Pattern Pair} in link $i$. The \emph{RSS mean} statistics $y_i(t)$ for link $i$ for dRTI is $y_i(t) = \sum_{j \in F_i} | R_{i,j}(t) - \bar{R}_{i,j}|$.


\item  \textbf{RSS variance}
The \emph{RSS variance} statistics for dRTI is computed over a window of $v$ measurements. The link quality statistics $y_i(t)$ for link $i$ for dRTI is $y_i(t) = \sum_{j \in F_i} var(R_{i,j}(t), ...., R_{i,j}(t - v + 1))$


\end{itemize}

\vspace{-2.5mm}
\section{Evaluation} \label{sec:experiments}

\vspace{-2.5mm}
\subsection{Goals, Metrics and Methodology}
\label{subsec:goals}
The goals of our evaluation are to study: 1) whether dRTI can lead to better tracking accuracy
in comparison to previous approaches based on omni-directional antennas and multi-channel RTI (cRTI); 
2) the performance of Location, Fade Level and PRR \emph{Pattern Pair} \bo{selection methods}; and 3) the energy overhead of dRTI.

We use two metrics to measure the tracking accuracy.
1) \emph{Root Mean Squared error (RMSE):} RMSE ($e_{rms}$) characterises the mean tracking error over the experiment and is defined as:
\begin{equation}
e_{rms} =  \sqrt{ \frac{1}{t_d-t_c} \sum_{t = t_c}^{t_d} { \hat{e} (t)^2}  } , \label{eqn:errrms}
\end{equation}
where, $t_c$ and $t_d$ are the experiment start time and end time respectively, and $\hat{e} (t)$ is the tracking error at time $t$, which is expressed as   
\begin{equation}
{ \hat{e}(t)} =   \parallel {\hat{x}(t)-x_g(t)} \parallel , \label{eqn:errrms1}
\end{equation}
where $\hat{x}(t)$ and $x_g(t)$ are the coordinates of the estimated location and ground truth at time $t$ respectively. 
2) \emph{Cumulative distribution function (CDF):}  CDF of an error level $\ell$ is the probability that the tracking error is less than or equal to $\ell$.
\bo{
3) \emph{Radio link attenuation false negative(FN) and false positive(FP):}  
 The definitions of these two qualities have already been discussed in Section~\ref{sec:hypoFNFP}. We will present FN and FP as a percentage. FN (FP) is the ratio of the number of FN (FP) links to the total number of links. 
}
\vspace{-2.5mm}

\subsubsection{Hardware and Software}
\label{subsubsec:hardware}
We deployed a network of seven nodes with ESD antennas and one base station, which collected the RSSI measurements from all the nodes and transferred 
them to a PC for RTI and tracking via a serial cable. We used the nodes, which have the same hardware (TelosB and ESD antenna with six directions) and operating system (Contiki) as those used in Section~\ref{subsubsec:setup}. TelosB nodes with CC2420 transceivers operate in the 2.4 GHz ISM band and the radio channels were picked from a list \{11, 15, 18, 21, 26\} as in~\cite{KaltiokallioBP12}.  For cRTI, we used 4 channels because it showed the best tracking performance~\cite{KaltiokallioBP12}.

The nodes run a simple token passing protocol in a TDMA 
fashion similar to SPIN\footnote{SPIN: TinyOS code for RSS collection. SPIN: TinyOS code for RSS collection
\url{http://span.ece.utah.edu/spin}} to produce and collect radio link RSSI measurements.
At a particular time, only one node transmitted packets, and the rest of the nodes received packets to measure pairwise RSS. 
The transmitter sent one packet per channel for cRTI and six packets per direction for dRTI in each round. The receivers \emph{dynamically} switched
antenna directions to try to collect one packet from each direction.

As introduced in Section~\ref{subsec:bgRTI}, if the value of a voxel in the RTI image has a larger value compared to the rest, it is
likely that an obstruction (a person) is located in this voxel. As in previous work~\cite{WilsonRTI:2010,WilsonVRTI:2011}, we
use the index of the voxel that has the largest value in the image as the estimated location of the person since 
we have one person only in the experiments. We further apply a Kalman filter to produce the
trajectory of the person. 
We use a voxel width of 0.2 metre, and $\lambda$ in Eq.~(\ref{eqn:rtiweight}) as 1.5 metres.
A webcam was deployed in AoI to record ground truth for tracking.

\vspace{-2.5mm}
\subsubsection{Experiment Description}
\begin{figure*}[th]
{
    \centering
       \subfigure[mRTI]{
        \includegraphics[scale = .6]{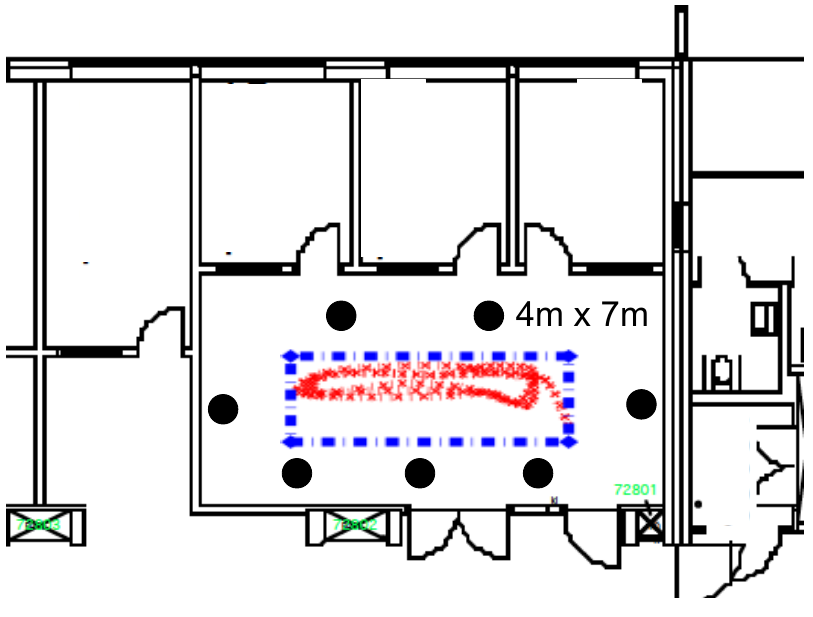}
        \label{fig:mRTILOS}
    }
    \subfigure[cRTI (RSS mean)]{
    \includegraphics[scale = .6]{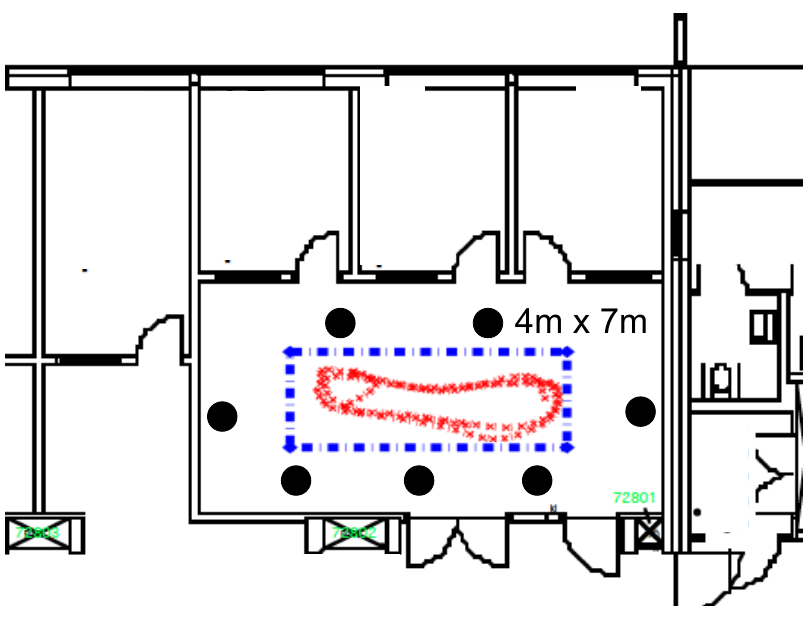}    
    \label{fig:cRTImLOS}
    }
     \subfigure[dRTI (RSS mean)]{
        \includegraphics[scale = .6]{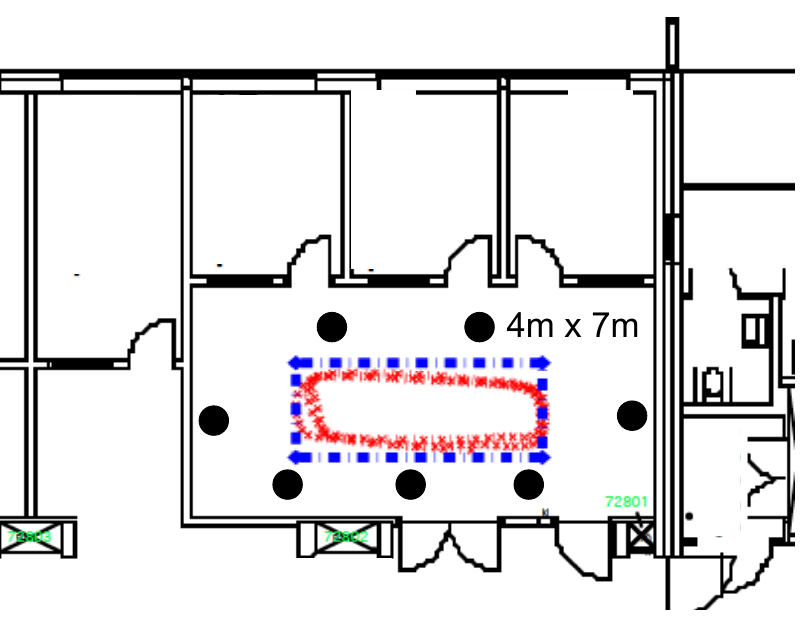}
        \label{fig:dRTImLOS}
    }
     \caption{The tracking performance of mean based RTI in the LOS experiment of omni RTI, cRTI, and dRTI. The red lines 
     are the trajectory estimation by different methods, and the dotted blue lines are the ground truth (\emph{Experiment 3}).}
           \label{fig:trackinglos_mean}
           }
       \vspace{-2.5mm}
\end{figure*}

We conducted 4 experiments in our lab. Experiments 1 and 3 were conducted in a large open room for testing in a LOS environment. The black dots in Fig.~\ref{fig:mRTILOS} show how the 7 nodes were placed in these experiments. Experiments 2 and 4 were conducted in a NLOS or ``through-wall'' environment. Seven nodes were placed in 4 adjacent rooms with 4 nodes in one room and 1 node each in the other 3 rooms, see Fig.~\ref{fig:mRTINLOS}. There was normal office equipment such as desk and computers in those 3 rooms with one node each. 
The walls between the rooms are made of either wood or glass.

Experiments 1 and 2 were designed to study the tracking performance of different \emph{Pattern Pair} selection methods in LOS and NLOS environments respectively. For the purpose of comparison, we also include the tracking performance of using all 36 \emph{Pattern Pairs}, which is termed as \emph{dRTI-All}. Experiments 3 and 4 aimed to compare the tracking accuracy of: dRTI (RSS mean and RSS variance), omni RTI (mRTI and vRTI) and cRTI (both mean and variance based).

\vspace{-2.5mm}
\subsection{Comparing Different \emph{Pattern Pair} Selection Methods}
\begin{figure}[ht]
    {    \centering
     \subfigure[mean based RTI.]{
        \includegraphics[scale = .23]{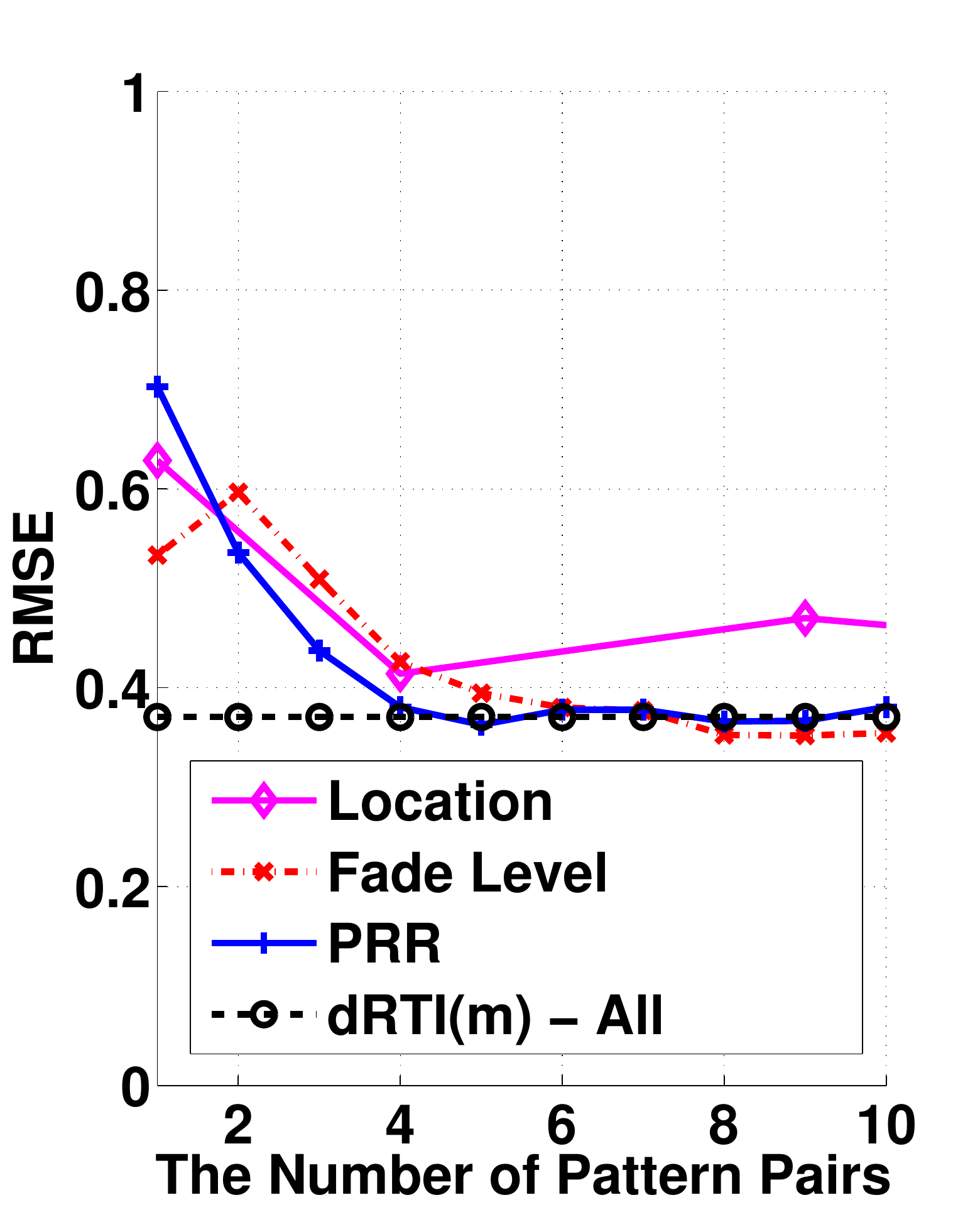}
        \label{fig:locationmeanlos}
    }
   \subfigure[variance based RTI]{
        \includegraphics[scale = .23]{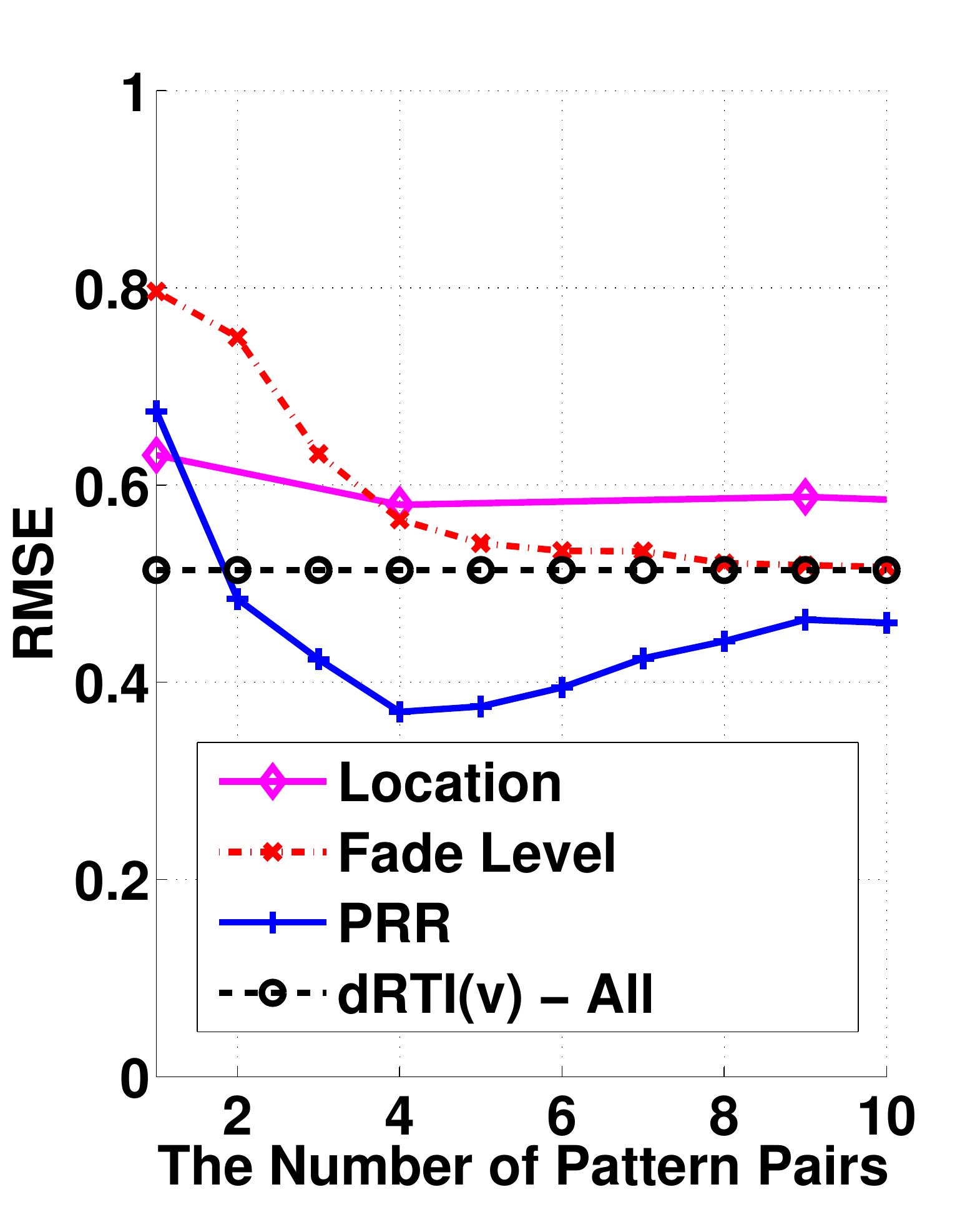}
        \label{fig:fadelevelmeanlos}
    }
        \caption{The tracking performance of \emph{Pattern Pair} selection methods in the indoor LOS environment (\emph{Experiment 1})}
        \label{fig:chosenpatternpair_los}}
          \vspace{-2.5mm}
\end{figure}

For both Experiments 1 (LOS) and 2 (NLOS), we apply the three \emph{Pattern Pair} selection methods to compare their performance. For each experiment, both mean and variance based statistics are used. For the mean based dRTI in the LOS environment, Fig.~\ref{fig:locationmeanlos} shows that the Fade Level and PRR methods perform similar to each other, and converge to the performance of dRTI-All when the number of \emph{Pattern Pairs} is larger than 5. For the variance based dRTI, Fig.~\ref{fig:fadelevelmeanlos} shows that the PRR method performs best and is even better than 
dRTI-All. The last result is not surprising because, from Fig.~\ref{fig:patternpairBoth}, we know that some \emph{Pattern Pairs} are more informative than others, and some \emph{Pattern Pairs} are even less informative than omni. These results also show that using all \emph{Pattern Pairs} is not necessarily better, and judicious selection can help improve performance and reduce overhead.
. 



\begin{figure}[ht]
      {  \centering
     \subfigure[mean based RTI.]{
        \includegraphics[scale = .23]{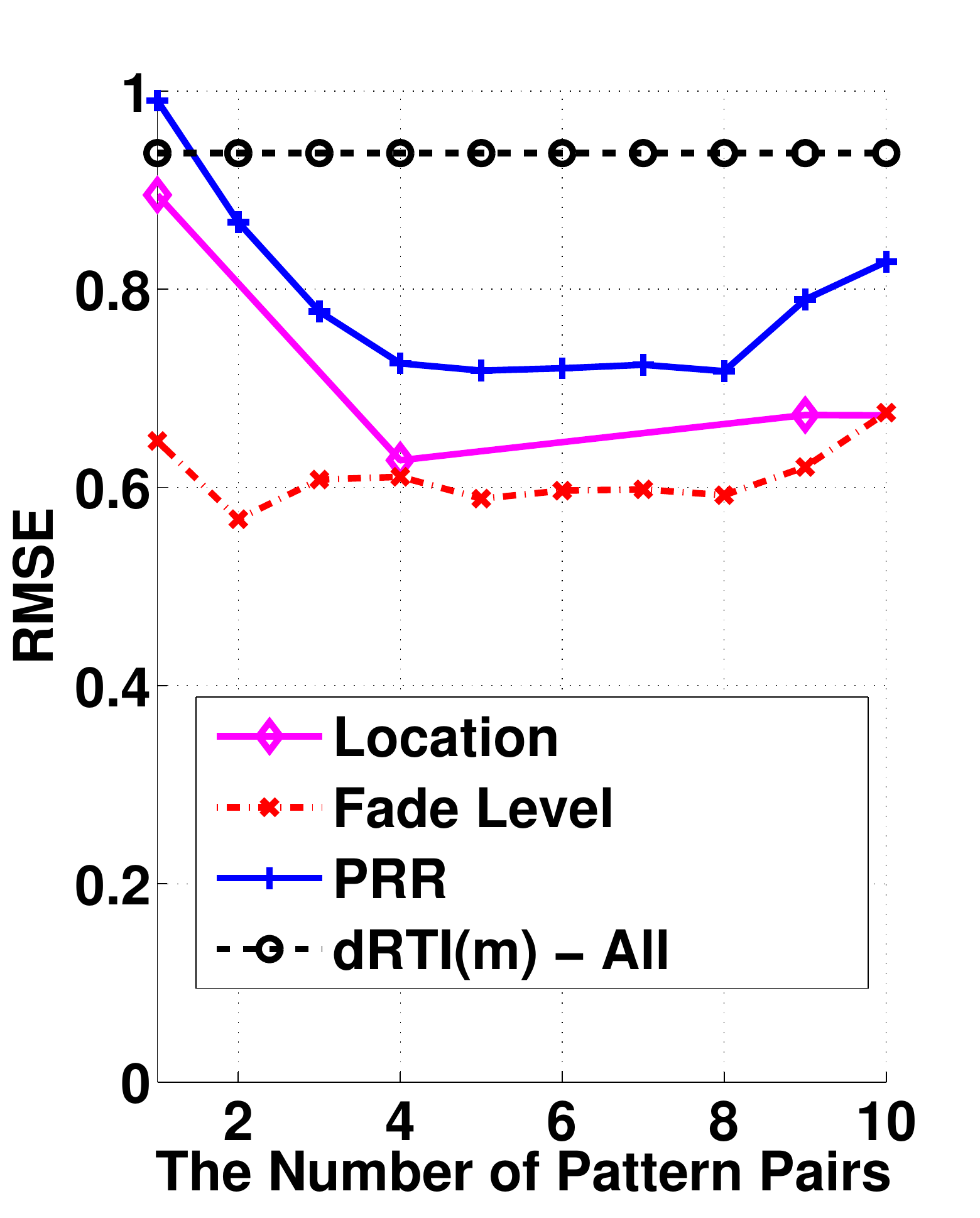}
        \label{fig:locationmeannlos}
    }
   \subfigure[variance based RTI]{
        \includegraphics[scale = .23]{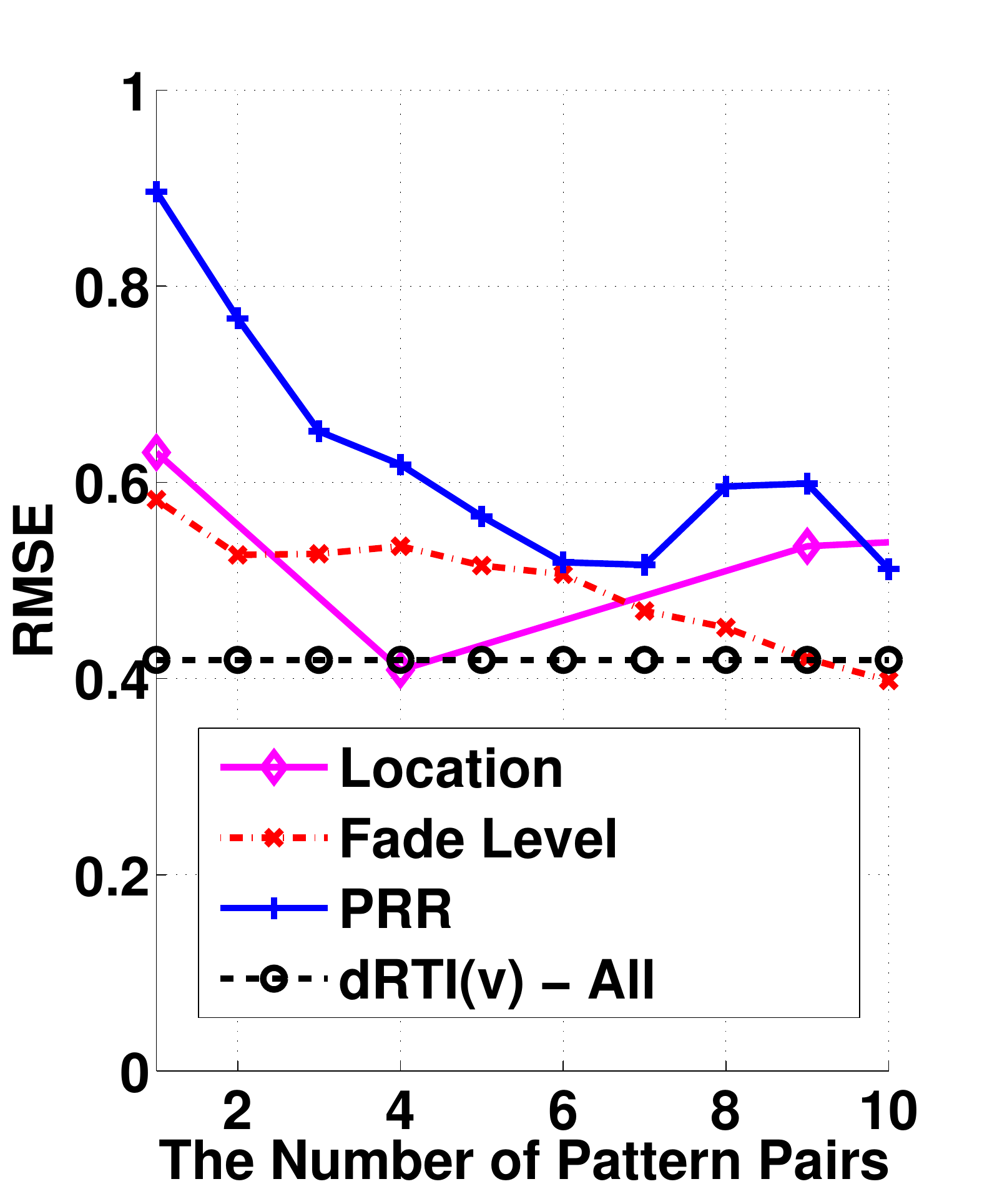}
        \label{fig:fadelevelmeannlos}
    }
    \caption{The tracking performance of \emph{Pattern Pair} selection methods in the indoor NLOS ``through-wall'' environment (\emph{Experiment 2})}
    \label{fig:chosenpatternpair_nlos}}
      \vspace{-2.5mm}
\end{figure} 

In the NLOS ``through-wall'' environment, variance based RTI (Fig.~\ref{fig:fadelevelmeannlos}) significantly outperforms mean based (Fig.~\ref{fig:locationmeannlos}) because the radio propagation environment is significantly more complicated than that of the LOS environment. Overall, Fig.~\ref{fig:chosenpatternpair_nlos}
shows that the Fade Level method is consistently better than the other methods. This is because the Fade Level method attempts to select the \emph{Pattern Pairs} that are \emph{least in deep fade} to provide the most information as discussed in Section~\ref{subsubsec:fade_level_method}.
Therefore, for Experiments 3 and 4, we will use the \emph{Fade Level} method for \emph{Pattern Pair} selection and use 9 \emph{Pattern Pairs}.


Fig.~\ref{fig:chosenpatternpair_los} and \ref{fig:chosenpatternpair_nlos}
 also show that the Location method  performs poorly compared to the other methods. Nevertheless, it can still provide reasonably good tracking performance (approximately $0.5$ metre) for both LOS (Fig.~\ref{fig:locationmeanlos}) and NLOS (Fig.~\ref{fig:fadelevelmeannlos})
environments. Furthermore, the Location method does not require calibration, which makes it very useful in emergency response applications.

\subsection{Comparing omni RTI, cRTI and dRTI}
\label{subsec:comprti}

Note that in this section we use mRTI and vRTI to refer to the mean-based and variance-based  \emph{omni} RTI because \bo{these} are names used in previous work. For dRTI and cRTI, it will be clear from the heading whether it is mean or variance based.

\subsubsection{\emph{Experiment 3}: LOS  environment} \label{subsubsec:los}

\vspace{-2.5mm}
\begin{table}[ht]
\footnotesize
   \centering
   \caption{$e_{rms}$ of mean based RTI in LOS experiment (\emph{Experiment 3}).}
    \begin{tabular}{ | l | l | l | l | }
\hline
method & mRTI  & cRTI & dRTI \\   \hline
$e_{rms} $(m) & $0.9054$  & $0.7872$ &$0.5221$ \\ 
  \hline
    \end{tabular}
      \label{tab:LOSerror_mean}
  \vspace{-2.5mm}
\end{table}

\begin{figure}[ht]
{
\centering
        \includegraphics[scale = .35]{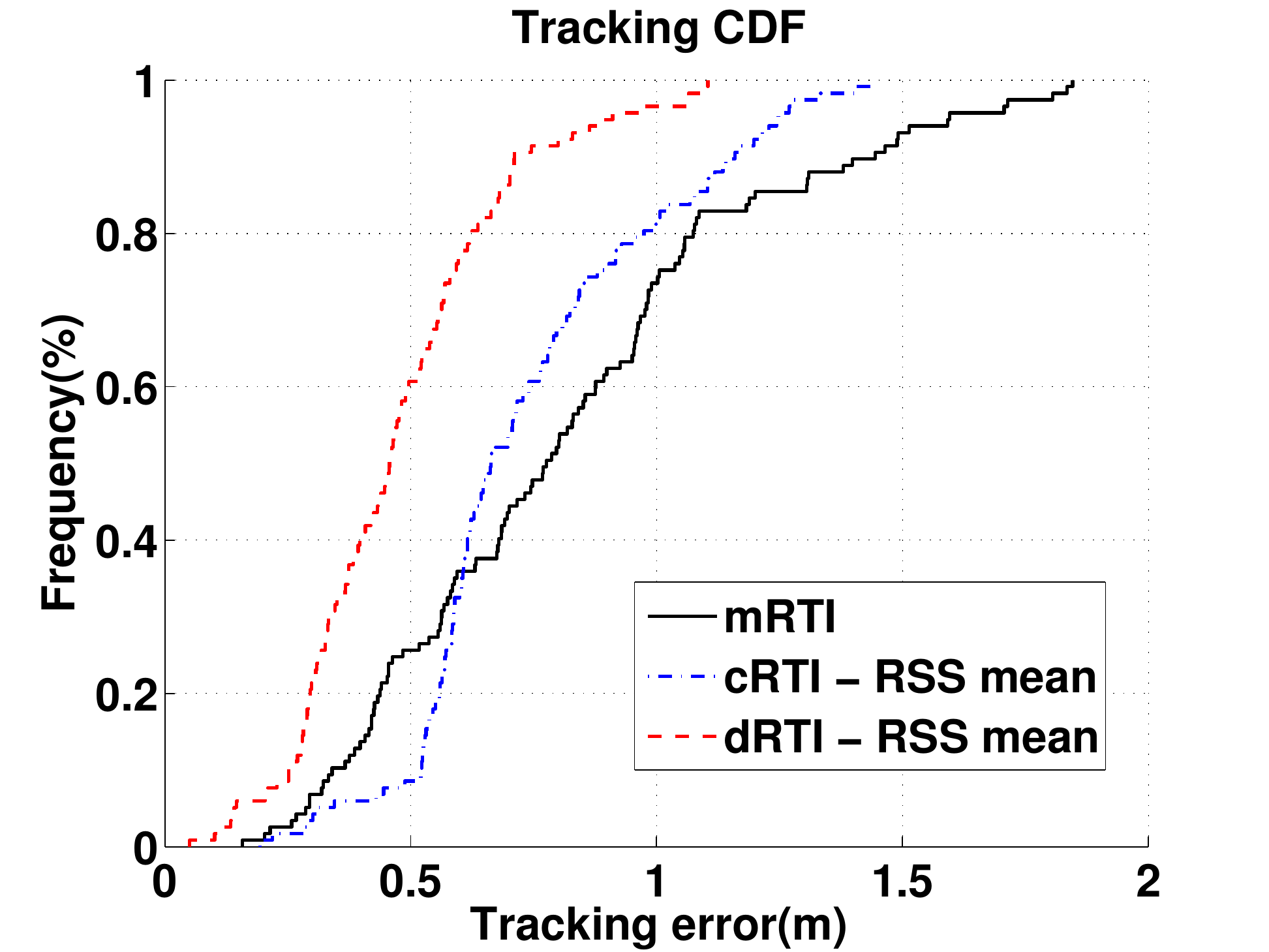}
        \caption{The CDFs of mean based RTI methods (\emph{Experiment 3}).}
            \label{fig:cdfmLOS} 
            }
    \vspace{-2.5mm}
\end{figure}

\textbf{Mean based methods:} Fig.~\ref{fig:trackinglos_mean} shows the tracking performance of mRTI, cRTI and dRTI in the indoor LOS environment. It depicts that the trajectory estimation 
of dRTI is significantly closer to the ground truth compared to both mRTI and cRTI. Table~\ref{tab:LOSerror_mean}
shows the $e_{rms}$ of mRTI, cRTI and dRTI are 0.9054 m, 0.7872 m, and 0.5221 m respectively.
dRTI achieved a tracking performance improvement of approximately 42\% (compared to mRTI) and 
approximately 34\% (compared to cRTI).  The CDFs for different
mean based RTI methods are presented in Fig.~\ref{fig:cdfmLOS}. The results show
that the 90th percentile tracking errors for mRTI, cRTI, and dRTI
were approximately 1.5 m, 1.2 m and 0.7 m respectively. dRTI achieved
performance improvements of approximately 53\% (compared to mRTI) 
and 42\% (compared to cRTI) respectively.

\begin{figure*}[ht]
{
    \centering
       \subfigure[vRTI]{
        \includegraphics[scale = .6]{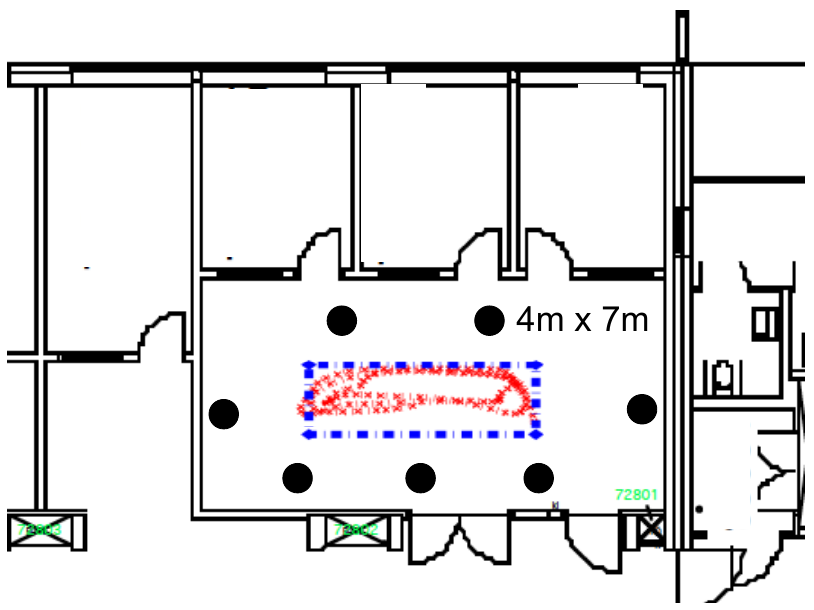}
        \label{fig:vRTILOS}
    }
    \subfigure[cRTI (RSS variance)]{
    \includegraphics[scale = .6]{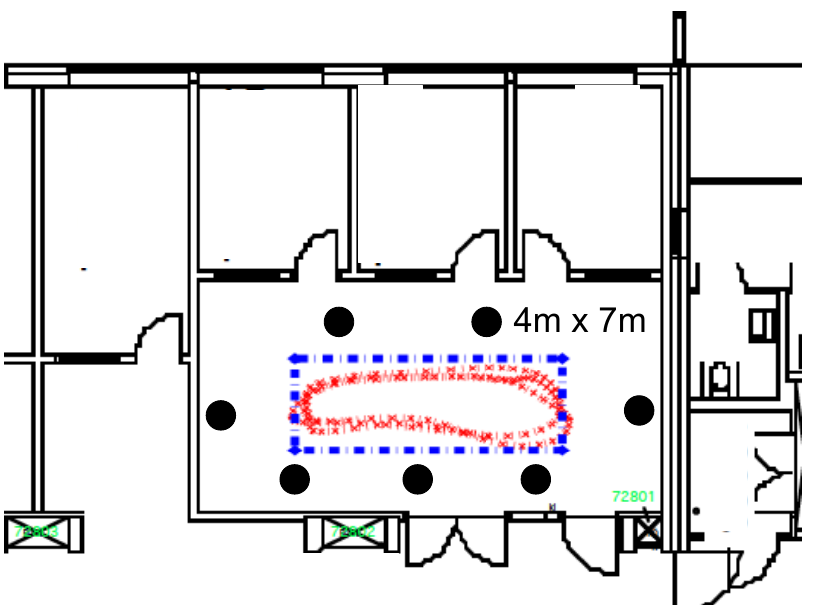}    
    \label{fig:cRTIvLOS}
    }
    \subfigure[dRTI (RSS variance)]{
        \includegraphics[scale = .6]{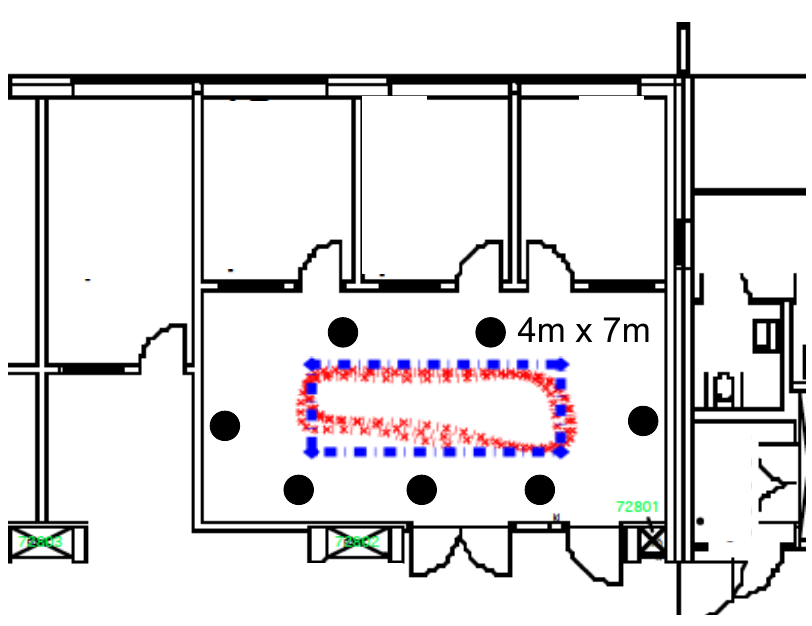}
        \label{fig:dRTIvLOS}
  }

    \caption{The tracking performance of variance based RTI in the LOS experiment of RTI, cRTI, and dRTI. The red lines 
     are the trajectory estimation by different methods, and the dotted blue lines are the ground truth (\emph{Experiment 3}).}
     \label{fig:trackinglos_variance}
     }
       \vspace{-2.5mm}
\end{figure*}

\begin{figure}[ht]
   { \centering
    \includegraphics[scale = .35]{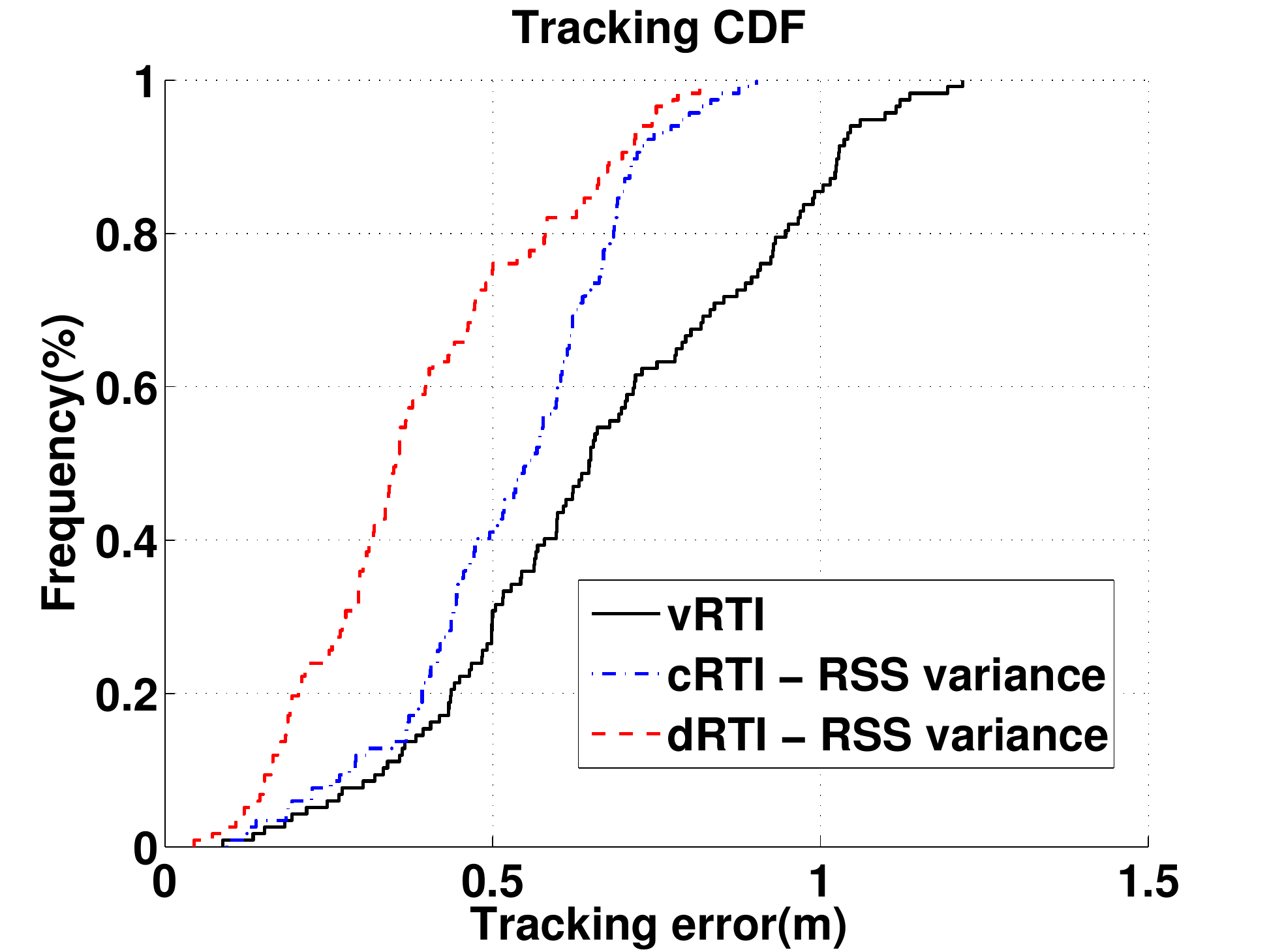} 
  \caption{The CDFs of variance based RTI methods (\emph{Experiment 3}).}
      \label{fig:cdfvLOS}
  }
    \vspace{-2.5mm}
\end{figure}

\textbf{Variance based methods:} Fig.~\ref{fig:trackinglos_variance} shows the tracking performance of vRTI, cRTI and dRTI demonstrating that the trajectory estimation 
of dRTI is significantly closer to the ground truth compared to both vRTI and cRTI.
Table~\ref{tab:LOSerror_variance}
shows, the $e_{rms}$ of vRTI, cRTI and dRTI are 0.7196 m, 0.5565 m, and 0.4340 m respectively.
dRTI achieved a tracking performance improvement of approximately 40\% (compared to vRTI) and 
approximately 22\% (compared to cRTI).  The CDFs for different
variance based RTI methods are presented in Fig.~\ref{fig:cdfvLOS}. The 90th percentile tracking errors for vRTI, cRTI, and dRTI
were approximately 1.1 m, 0.75 m and 0.7 m respectively. dRTI achieved
performance improvements of approximately 36\%  (compared to vRTI) 
and 7\% (compared to cRTI) respectively.

\begin{table}[ht]
\footnotesize
   \centering
   \caption{$e_{rms}$ of  variance based RTI in LOS experiment (\emph{Experiment 3}).}
    \begin{tabular}{ | l | l | l | l | }
\hline
method & vRTI  & cRTI & dRTI \\   \hline 
$e_{rms} $\bo{(m)} & $0.7196$ & $0.5565$  & $0.4340$ \\
 \hline
    \end{tabular}
\label{tab:LOSerror_variance}
  \vspace{-2.5mm}
\end{table}


\begin{figure}[ht]
    {    \centering
     \subfigure[mean based RTI.]{
        \includegraphics[scale = .35]{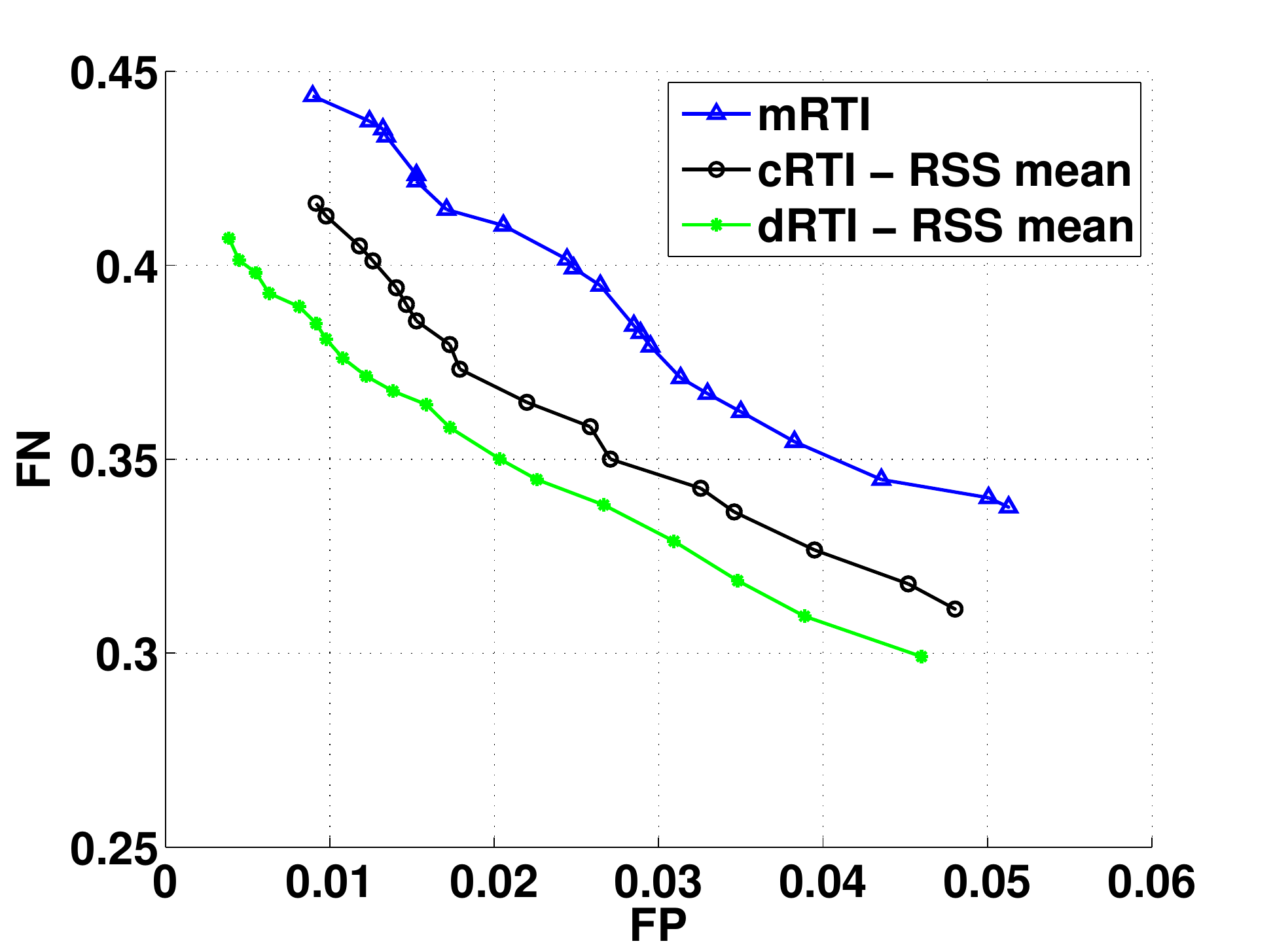}
        \label{fig:FNFPmeanlos}
    }
   \subfigure[variance based RTI]{
        \includegraphics[scale = .35]{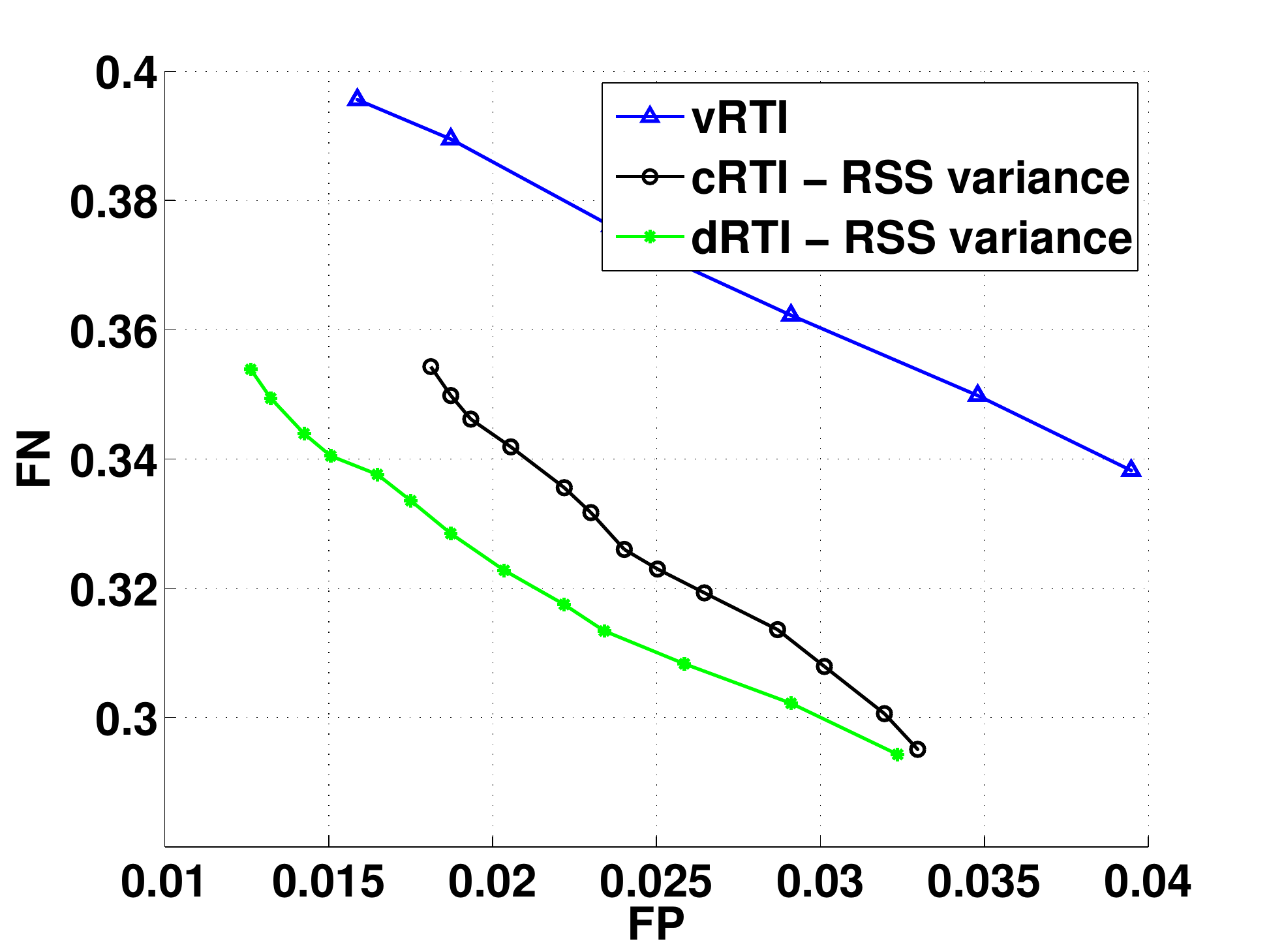}
        \label{fig:FNFPvarlos}
    }
        \caption{The statistics for Radio Link Attenuation FN and FP in the indoor LOS environment (\emph{Experiment 3})}
        \label{fig:FNFP_los}}
          \vspace{-2.5mm}
\end{figure} 
\textbf{Radio Attenuation FN and FP:} 
\bo{
Fig.~\ref{fig:FNFP_los} shows the statistics of radio link attenuation FN and FP. 
 The figure compares three classes of methods: m/vRTI,cRTI and dRTI. For each method, we vary the threshold used to determine whether a link is FP and FN. We know from the standard theory on statistical hypothesis testing that there is a trade-off between FP and FN. This trade-off is clearly visible in the figure. We know that if a method can achieve lower FP and FN, then it is a better method. It is clear from Fig.~\ref{fig:FNFP_los} that dRTI has the lowest FP and FN compared with the other two methods. 
 These statistics support the hypotheses in Section \ref{sec:hypoFNFP} and explain the reason why dRTI perform better than omni-directional RTI and cRTI. 
}


\vspace{-2.5mm}
\subsubsection{\emph{Experiment 4}:  NLOS ``Through-wall'' environment} 
\label{subsubsec:nlos}

\begin{table}[ht]
\footnotesize
   \centering
   \caption{$e_{rms}$ of mean based RTI in NLOS experiment (\emph{Experiment 4}).}
    \begin{tabular}{ | l | l | l | l | }
\hline
method & mRTI  & cRTI & dRTI \\   \hline 
$e_{rms} $\bo{(m)} & $1.4922 $  & $0.8580 $ & $0.7506 $ \\   
 \hline
    \end{tabular}
\label{tab:NLOSerror_mean}
  \vspace{-2.5mm}
\end{table}

\begin{figure*}[ht]
{
    \centering
       \subfigure[mRTI]{
        \includegraphics[scale = .60]{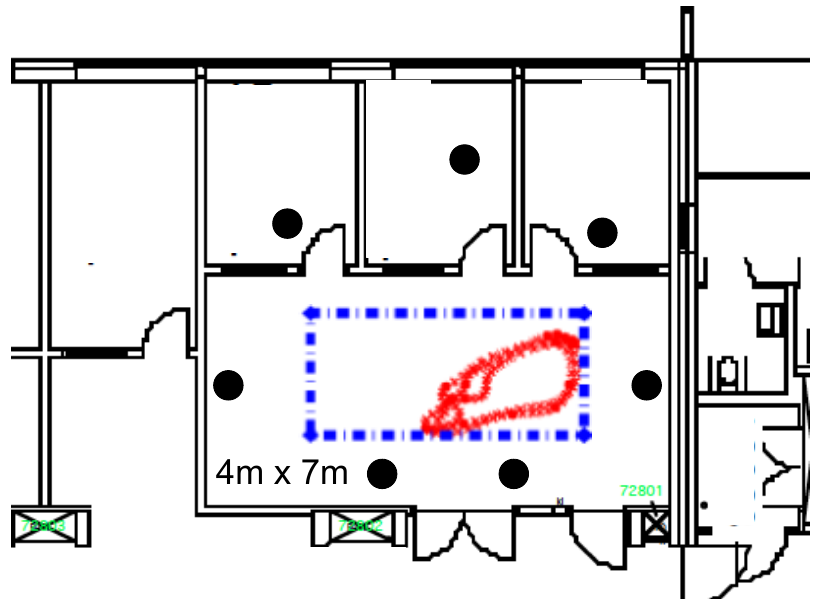}
        \label{fig:mRTINLOS}
    }
    \subfigure[cRTI (RSS mean)]{
    \includegraphics[scale = .60]{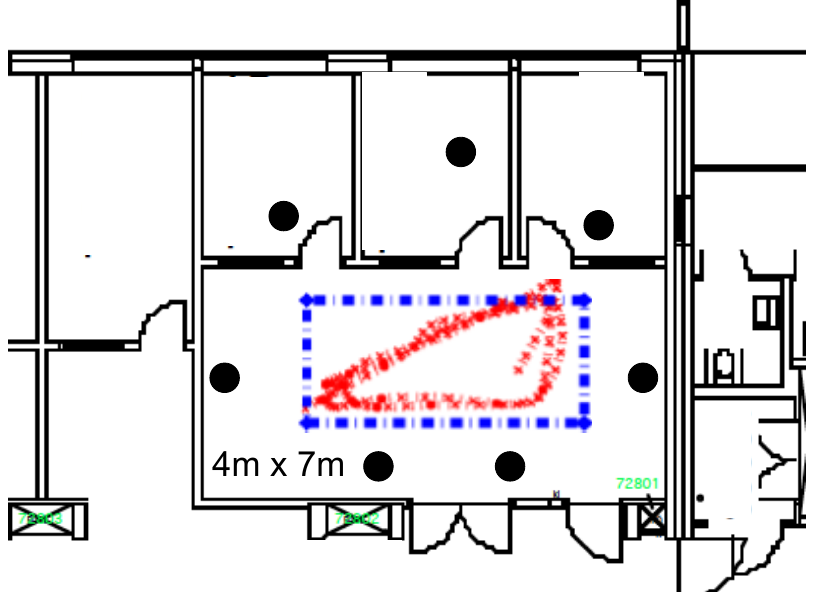}    
    \label{fig:cRTImNLOS}
    }
     \subfigure[dRTI (RSS mean)]{
        \includegraphics[scale = .60]{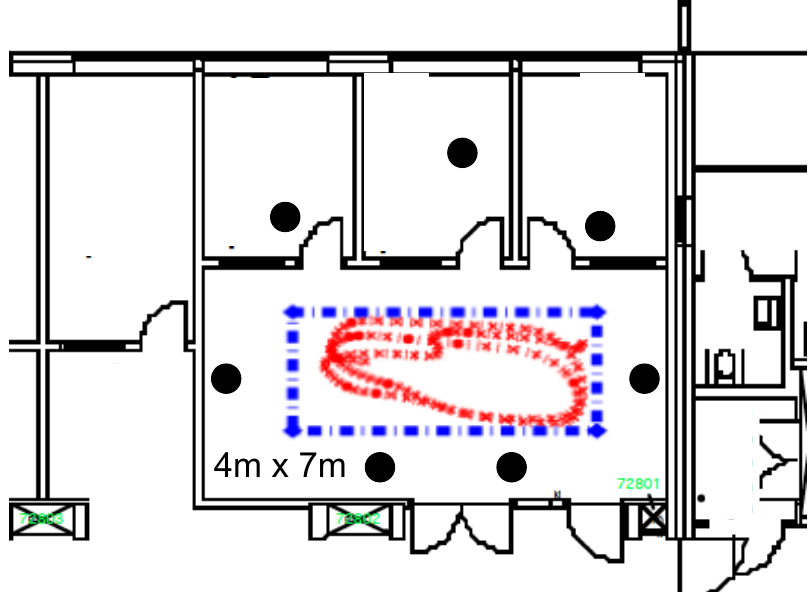}
        \label{fig:dRTImNLOS}
          \vspace{-2.5mm}
    }
     \caption{The tracking performance of mean based RTI in the NLOS ``through-wall'' experiment of RTI, cRTI, and dRTI. The red lines 
     are the trajectory estimation by different methods, and the dotted blue lines are the ground truth (\emph{Experiment 4}).}
          \label{fig:trackingwall_mean}}
\end{figure*}

\begin{figure}[ht]
   { \centering
        \includegraphics[scale = .35]{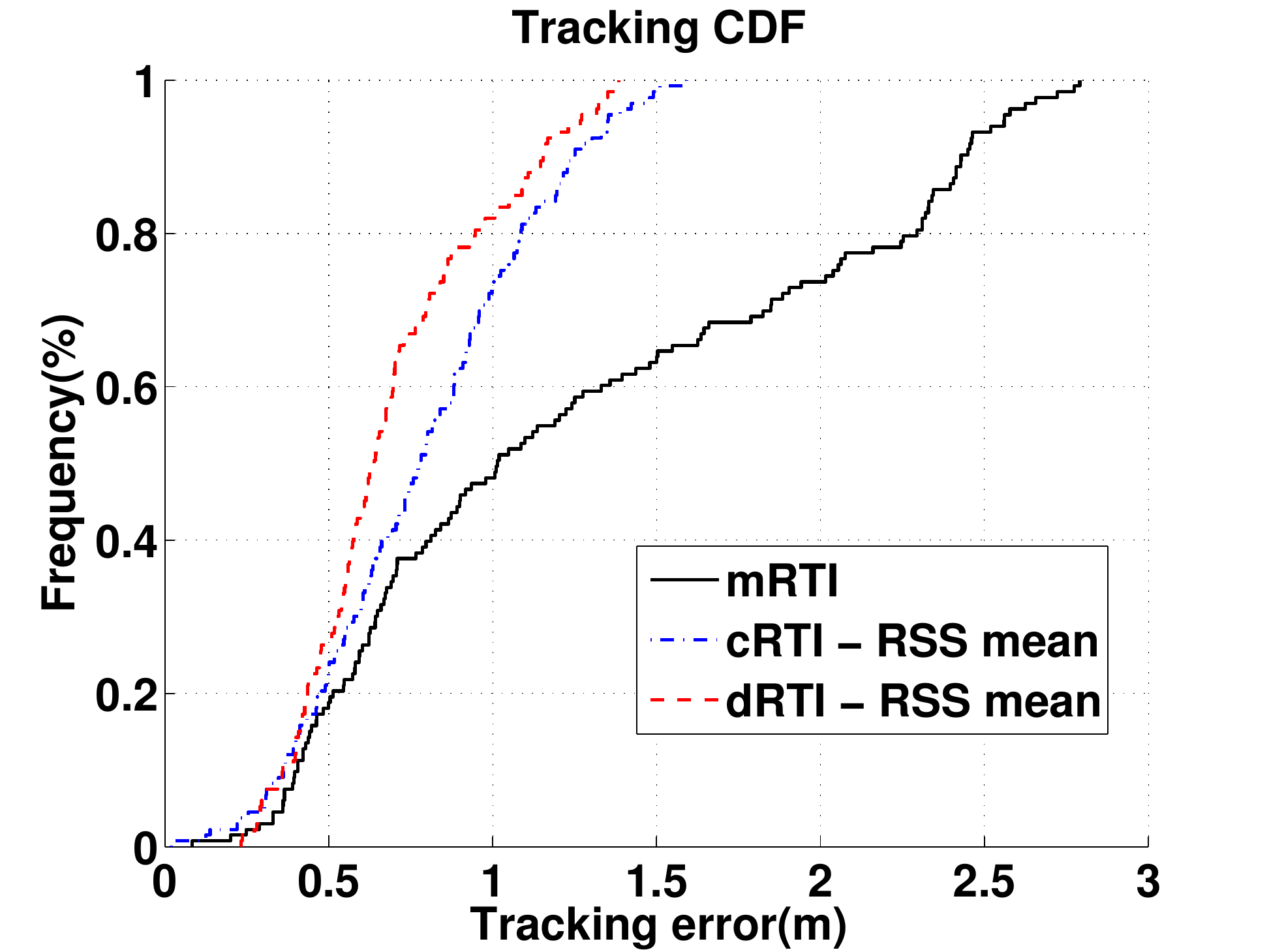}
        \label{fig:cdfmNLOS}
    \label{fig:cdfNLOS}
  \caption{The CDFs of mean based RTI methods (\emph{Experiment 4}).}
  \label{fig:cdf_mean_4}
  }
    \vspace{-2.5mm}
\end{figure}

\textbf{Mean based methods:} Compared to the performance in the LOS environment in Section~\ref{subsubsec:los},
the NLOS ``through-wall'' environment is significantly more challenging for RTI, and the
tracking errors are significantly higher for all RTI methods.
However, Fig.~\ref{fig:trackingwall_mean} shows the tracking performance of 
mRTI, cRTI and dRTI in a NLOS ``through-wall'' environment. It demonstrates that the trajectory estimation 
of dRTI is significantly closer to the ground truth compared to both mRTI and cRTI. Table~\ref{tab:NLOSerror_mean}
shows, the $e_{rms}$ of mRTI, cRTI and dRTI are 1.4922 m, 0.8580 m, and 0.7506 m respectively.
dRTI achieves a tracking performance improvement of approximately 50\% (compared to mRTI) and 
approximately 12\% (compared to cRTI).  The CDFs for different
mean based RTI methods were showed in Fig.~\ref{fig:cdf_mean_4}. The 90th percentile tracking errors for mRTI, cRTI, and dRTI
were  approximately 2.4 m, 1.3 m and 1.2 m respectively. dRTI achieved
performance improvements of approximately 50\% (compared to mRTI) 
and 8\% (compared to cRTI) respectively.


\begin{table}[ht]
\footnotesize
   \centering
   \caption{$e_{rms}$ of variance based RTI in NLOS experiment (\emph{Experiment 4}).}
     \begin{tabular}{ | l | l | l | l | }
\hline
method & vRTI  & cRTI & dRTI \\   \hline 
$e_{rms} $\bo{(m)} & $1.0432 $ & $0.8574 $ & $0.6734 $ \\ 
\hline
    \end{tabular}
    \label{tab:NLOSerror_variance}
  \vspace{-2.5mm}
\end{table}

\begin{figure*}[ht]
    {\centering
       \subfigure[vRTI]{
        \includegraphics[scale = .6]{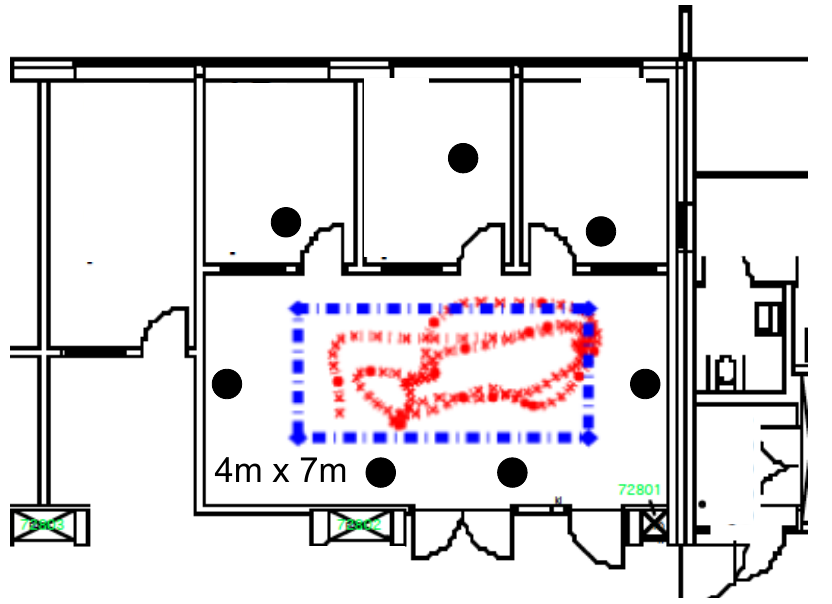}
        \label{fig:vRTINLOS}
    }
    \subfigure[cRTI(RSS variance)]{
    \includegraphics[scale = .6]{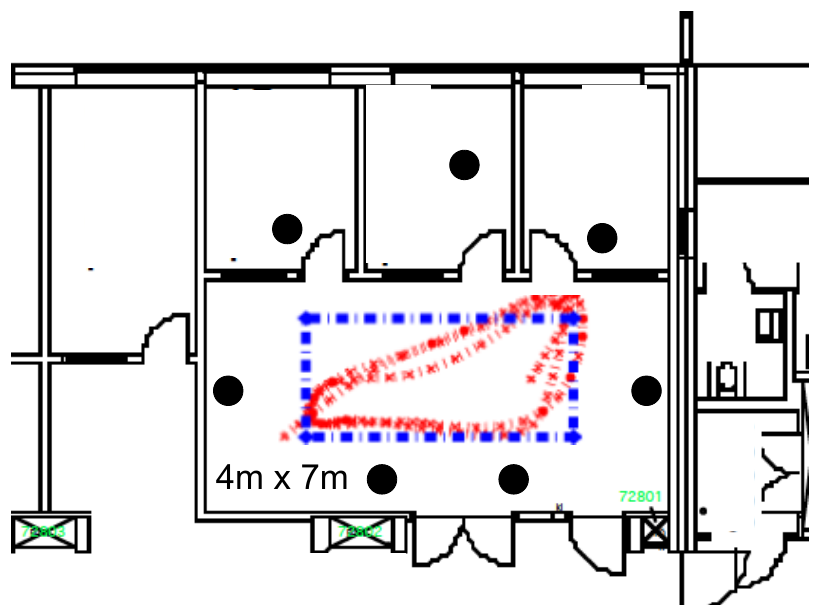}    
    \label{fig:cRTIvNLOS}
    }
            \subfigure[dRTI(RSS variance)]{
        \includegraphics[scale = .6]{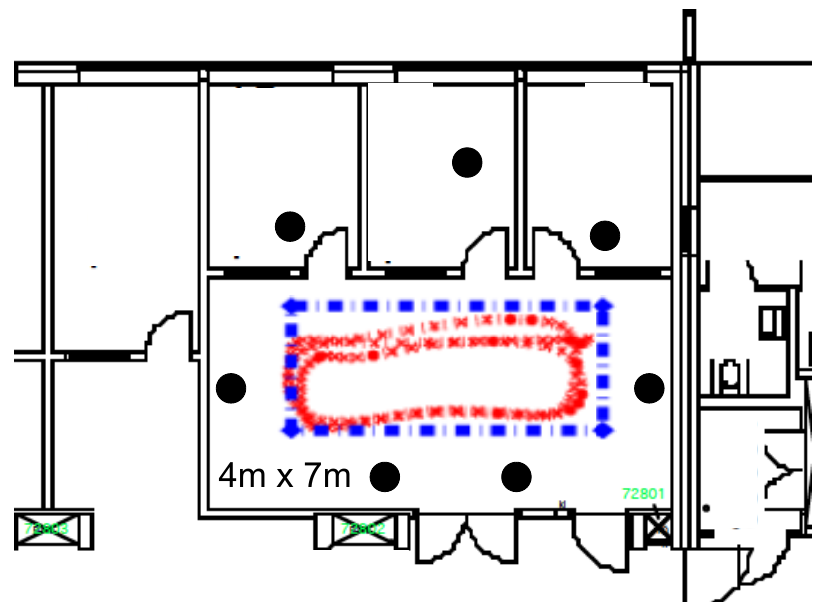}
        \label{fig:dRTIvNLOS}
  }
    \caption{The tracking performance of variance based RTI in NLOS ``through-wall'' experiment of RTI, cRTI, and dRTI. The red lines 
     are the trajectory estimation by different methods, and the dotted blue lines are ground true (\emph{Experiment 4}).}
    \label{fig:trackingwall_variance}}
    \vspace{-2.5mm}
\end{figure*}

\begin{figure}[ht]
{\centering    
    \includegraphics[scale = .35]{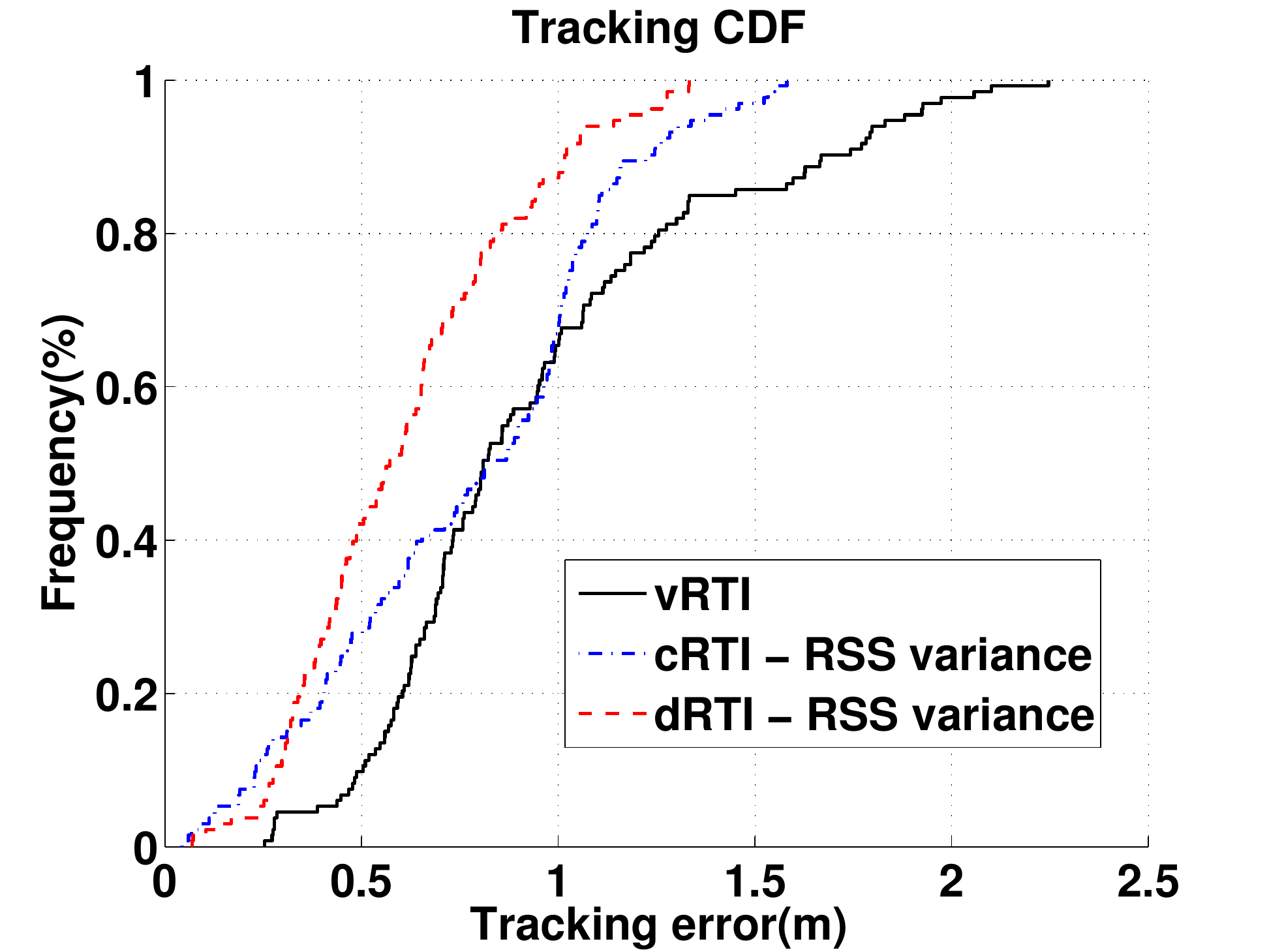} 
    \label{fig:cdfvNLOS}
     \caption{The CDFs of variance based RTI methods (\emph{Experiment 4}).}\label{fig:cdf_variance_4}
    }
    \vspace{-2.5mm}
\end{figure}

\textbf{Variance based methods:} Previous work showed that variance based RTI methods significantly outperformed
mean based RTI methods in NLOS ``through-wall'' environments~\cite{WilsonVRTI:2011}.
Our results in Tables~\ref{tab:NLOSerror_mean} and \ref{tab:NLOSerror_variance} also confirm this finding.
Furthermore, Fig.~\ref{fig:trackingwall_variance} shows that the trajectory estimation 
of dRTI is significantly closer to the ground truth compared to both vRTI and cRTI. Table~\ref{tab:NLOSerror_variance}
also
shows, the $e_{rms}$ of vRTI, cRTI and dRTI are 1.0432 m, 0.8574 m, and 0.6734 m respectively.
dRTI achieves a tracking performance improvement of approximately 35\% (compared to vRTI) and 
approximately 21\% (compared to cRTI).  The CDFs for different
variance based RTI methods are presented in Fig.~\ref{fig:cdf_variance_4}. The 90th percent of tracking errors for vRTI, cRTI, and dRTI
were  approximately 1.7 m, 1.2 m and 1 m respectively. dRTI achieves
performance improvements of approximately 41\% (compared to vRTI) 
and 17\% (compared to cRTI) respectively.

\begin{figure}[ht]
      {  \centering
     \subfigure[mean based RTI.]{
        \includegraphics[scale = .35]{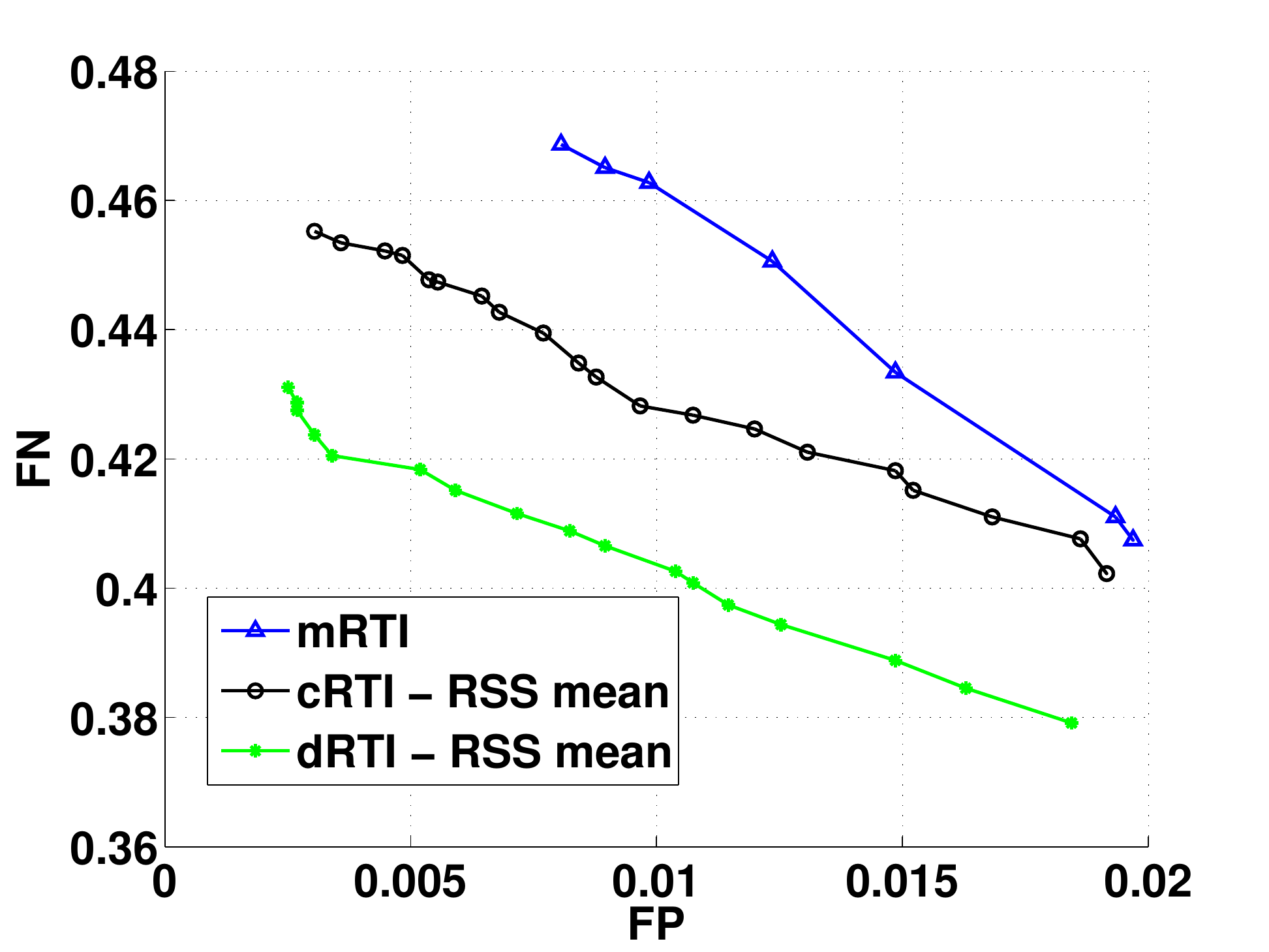}
        \label{fig:FNFPmeannlos}
    }
   \subfigure[variance based RTI]{
        \includegraphics[scale = .35]{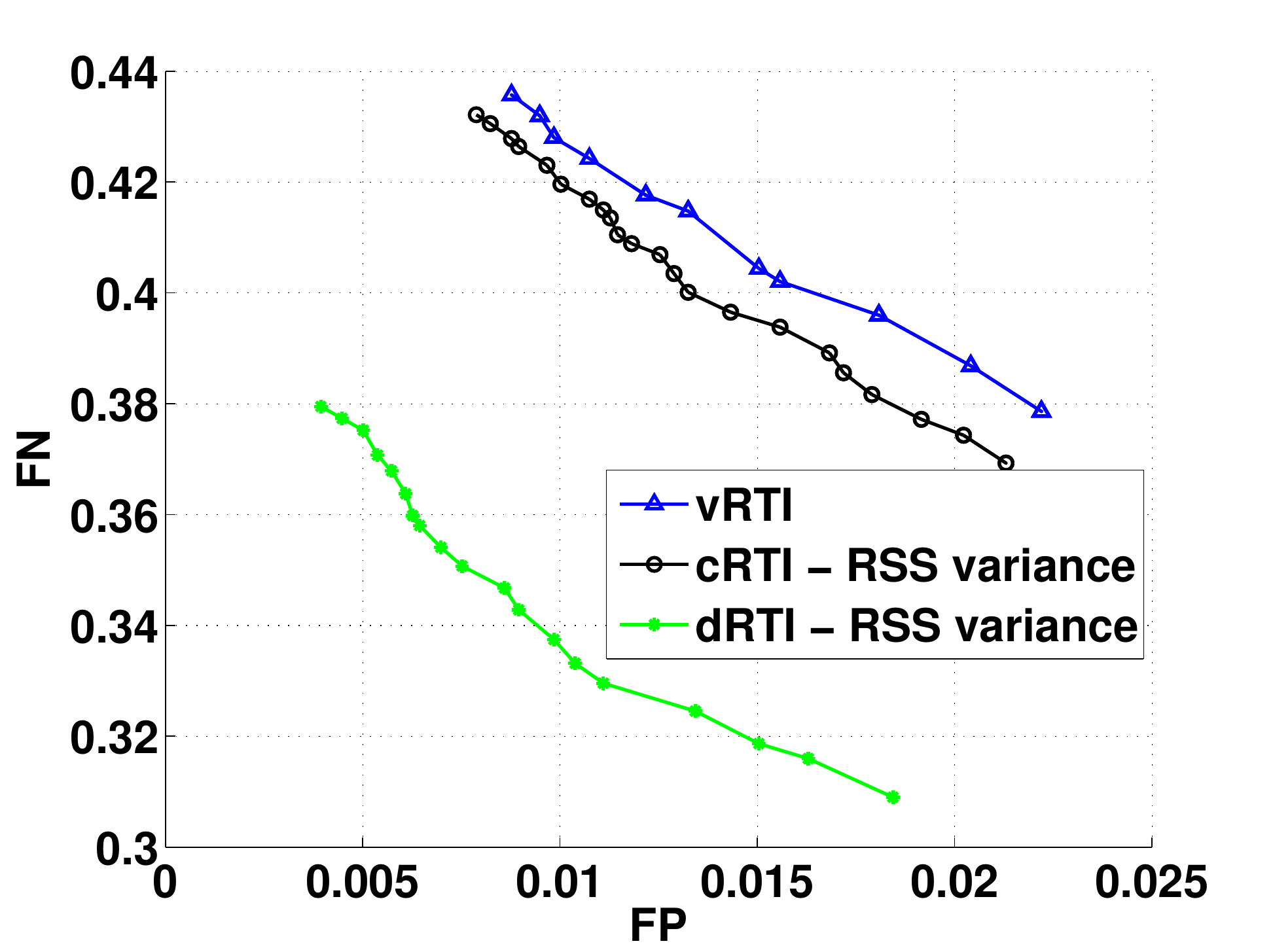}
        \label{fig:FNFPvarnlos}
    }
    \caption{The statistics for Radio Link Attenuation FN and FP  in the indoor NLOS ``through-wall'' environment (\emph{Experiment 4})}
    \label{fig:FNFP_nlos}}
      \vspace{-2.5mm}
\end{figure} 

\textbf{Radio Attenuation FN and FP:} 
\bo{
Fig.~\ref{fig:FNFP_nlos} shows similar results compared to those discussed in Section~\ref{subsubsec:los}. The figure shows that dRTI has the lowest number of FP and FN compare to those of omni-directional RTI and cRTI. 
}

\vspace{-2.5mm}
\subsection{Energy Cost}
We measured the energy consumption of our dRTI system by connecting a node to a 10 ohm resistor in series of the power supply and measured the voltage across the resistor by an oscilloscope. 
For small beacon packets whose payload length is 5 bytes, consisting of node ID (2 bytes), transmit direction (1 byte), and sequence number (2 bytes), the energy consumption was approximately 55 $\mu$J. For large packets for transferring RSS measurements to the base station with payload length of 62 bytes, the energy consumption was approximately 179 $\mu$J.
 The transmission time for the 5-byte beacon packets and 62-byte large packets were approximately 0.72 ms and 2.85 ms respectively.
The energy consumption was measured when the transmit power is set to the maximum value of 31.

\vspace{-2mm}
\section{Related work}\label{sec:related}
\vspace{-2mm}
\subsection{RF-Based Device Free Localisation}

Ultra-wideband (UWB) impulse radar devices have been studied for through-wall imaging. They transmit wideband impulses, and estimate the distance to a scatterer by measuring the time delay of the echo. %
UWB impulse radar typically performs well at short range. However, the accuracy decreases at long range because of limited transmit power and $d^4$ scattering losses \cite{WilsonRTI:2010}. 

Recently, many RSS-based device-free localization methods have been proposed.  Youssef et al.\ used the variance or change in RSS to detect a person crossing a link~\cite{Youssef:2007}.  Patwari and Agrawal proposed to use tomography to estimate an image of the AoI based on the signal attenuation of many radio links (mRTI)~\cite{PatwariEffect2008}, which they called RTI. However, their following work in~\cite{WilsonVRTI:2011} showed that mRTI was ineffective in the ``through-wall'' setting  because the radio signal statistics behaved unexpectedly. Their experiment showed the RSS 
measurements may decrease, keep stable or even increase when a link was blocked by an obstruction. Therefore, instead of using mRTI, they proposed to use the variance-based RTI (vRTI) in~\cite{WilsonVRTI:2011}. Later, Kaltiokallio et al.\ proposed to use multiple independent channels to increase the performance of 
RTI (cRTI) in~\cite{KaltiokallioBP12}. In cRTI, performance improves with the number of channels measured. Channel diversity requires more bandwidth compared to single-channel RTI, and is limited by the number of ``clean'' or interference-free radio channels.  \emph{With ubiquitous use of wireless technologies such as WiFi, it becomes more and more difficult to dedicate bandwidth, particularly in ISM bands, for RTI systems.} To this end, the proposed dRTI requires only one channel, the same as the traditional single-channel RTI.

RSS fingerprint-based methods for device free indoor localisation and tracking have been investigated in~\cite{Xu:2012,Xu:2013}.  Fingerprint-based methods do not require the sensor coordinates to be known, and they adaptively learn the multipath characteristics of an environment.  This paper shows that use of directional antennas in a wireless network increases the effect that a person has on the RSS.  Further, having a diversity of directional patterns effectively increase the dimension of the RSS fingerprint.  Both effects could be applied to improve the accuracy of RSS fingerprint-based methods as well, although the focus in this paper has been on RTI. 

Fingerprint-based methods require training periods in which a person is required to walk in each voxel in the AoI while the systems collect RSS fingerprints, as well as an empty-room training period. In contrast, mRTI requires only the empty-room training period, and vRTI requires no training period.  vRTI is advantageous since training is not possible in emergency response applications, however, vRTI is cannot detect static objects in the AoI.   Edelstein and Rabbat proposed a method to identify the background (no person obstructing) RSS from a radio link that has periods of obstruction and no obstruction \cite{10.1109/TMC.2012.206}. This work is complementary to the approaches in~\cite{10.1109/TMC.2012.206,ZhaoKRTI:2013} and could add directionality to improve their performance.  

Adib and Katabi recently proposed WiVi to count the number of people ``through-wall'' and recognise simple gestures based on WiFi signals in~\cite{Adib:2013:STW:2486001.2486039}.  WiVi can count the number of people and estimate their relative velocity.  Pu et al.\ developed WiSee, which demonstrated high classification accuracy among a set of nine gestures, using Doppler shift measurements and machine learning \cite{pu2013whole}.  In contrast, this paper focuses on tracking the absolute positions of people indoors.

\vspace{-2mm}
\subsection{Directional Communications}
Directional communications have been studied in both WiFi networks and WSNs. 
The challenges imposed by directional communications at the MAC layer have been extensively studied over the past few years \cite{1348118,4359017,832169,Choudhury:2002:UDA:570645.570653,Takai:2002:DVC:513800.513823}. 
While the majority of the work is in outdoor environments, recent works showed that directional communications is useful in indoor environments as well.
Blanco et al.  observed that directional antennas can lead to performance improvement over omni in an indoor environment and made an important observation that LOS direction did not always have the strongest signal strength in~\cite{Blanco:2008:ESB:1791949.1791967}.  
Liu et al. suggested that solutions developed for outdoor environments with directional communications might not work well in indoor environments, but demonstrated improved network capacity on the usage of directional communications~\cite{Liu:2009:DII:1592568.1592589, Liu:2010:PEI:1859995.1860020}. Lakshmanan et al.~highlighted
 that the benefits of directional communications are mostly resulted from the link gains instead of spatial medium reuse  in indoor environments \bo{\cite{Lakshmanan2010}}. 

In WSNs, Giorgetti et al.~demonstrated that a four-beam patch antenna could increase transmissions ranges and decrease signal variability and interference suppression from WiFi networks in~\cite{giorgetti07directional}. Voigt et al.~investigated indoor link dynamics for
directional communications and observed that the best direction for communications depended on the wireless channels used~\cite{thiemo}.  
Varshney et al. demonstrated that, in an indoor environment, directional transmissions and receptions together could significantly alleviate wireless contentions, which allowed better usage of the available wireless channels in~\cite{AmbujSPIDA}. 

To the best of our knowledge, this paper is the first to study the feasibility and performance of RTI for localisation and tracking with directional communications. 

\vspace{-2mm}
\section{Conclusions}\label{sec:conclusion}
In this paper, we investigate how directional antennas can be used to improve localisation accuracy of RTI. We implement dRTI and show that obstructions have a greater impact on the link quality of directional links. This in turn significantly improves the localisation accuracy of dRTI. Since the number of antenna direction pairs between two dRTI nodes can be large (e.g., 36 in our dRTI system), we further propose methods to effectively select informative antenna direction pairs to reduce communication and energy overhead. We evaluate the proposed dRTI system in different indoor environments. Our extensive experiments show that dRTI significantly outperforms omni-directional RTI and multi-channel RTI
in both line-of-sight (LOS) and non-LOS ``through-wall'' environments. 

\bibliographystyle{abbrv}
\small{
\bibliography{sigproc}  
}
%
%


\end{document}